\documentclass{jfm}
\usepackage{graphicx}
\usepackage{epstopdf, epsfig}
\usepackage[hidelinks]{hyperref}

\usepackage{color}
\newcommand{\makered}[1]{\textcolor{black}{#1}} 
\usepackage{color}
\newcommand{\makeredTwo}[1]{\textcolor{black}{#1}} 

\shorttitle{Bubble--particle collisions in turbulence}
\shortauthor{T. T. K. Chan, C. S. Ng and D. Krug}

\title{Bubble--particle collisions in turbulence: insights from point-particle simulations}

\author{Timothy T. K. Chan\aff{1}\corresp{\email{t.k.t.chan@utwente.nl}},
	Chong Shen Ng\aff{1}
	\and Dominik Krug\aff{1}\corresp{\email{d.j.krug@utwente.nl}}}

\affiliation{\aff{1}Physics of Fluids Group, Max Planck Center Twente for Complex Fluid Dynamics, Faculty of Science and Technology, MESA+ Research Institute, and J. M. Burgers Centre for Fluids Dynamics, University of Twente, P.O. Box 217, 7500 AE Enschede, The Netherlands}

\begin{document}
	
	\maketitle
	
	\begin{abstract}
		Bubble--particle collisions in turbulence are central to a variety of processes such as froth flotation. Despite their importance, details of the collision process have not received much attention yet. This is compounded by the sometimes counter-intuitive behaviour of bubbles and particles in turbulence, as exemplified by the fact that they segregate in space. Although bubble--particle relative behaviour is fundamentally different from that of identical particles, the existing theoretical models are nearly all extensions of theories for particle--particle collisions in turbulence. The adequacy of these theories has yet to be assessed as appropriate data remain scarce to date. \makered{In this investigation, we study the geometric collision rate by means of} direct numerical simulations of bubble--particle collisions in homogeneous isotropic turbulence using the point-particle approach over a range of the relevant parameters, including the Stokes and Reynolds numbers. We analyse the spatial distribution of bubble and particles, and quantify to what extent their segregation reduces the collision rate. This effect is countered by increased approach velocities for bubble--particle compared to monodisperse pairs, which we relate to the difference in how bubbles and particles respond to fluid accelerations. \makered{We found that in the investigated parameter range, these collision statistics are not altered significantly by the inclusion of a lift force or different drag parametrisations, or when assuming infinite particle density.} Furthermore, we critically examine existing models and discuss inconsistencies therein that contribute to the discrepancy. 
	\end{abstract}
	
	
	\section{Introduction}\label{sec::intro}
	Collisions between bubbles and particles in a turbulent flow are of significant technological relevance. In particular, such collisions are essential to the flotation process, which is a widely used separation technique especially in the mining industry \citep{nguyen_colloidal_2004}. In this process, after grinding, small ore fragments are fed into a big water-filled cell that is agitated via a rotor and into which air bubbles are injected. Collisions between the ore particles and the bubbles then form the base for the decisive test: valuable mineral particles attach to the bubbles due to their hydrophobic surface and consequently rise to the top where they can be skimmed off as a froth, whereas the hydrophilic waste rock particles remain in suspension and are eventually discharged as tailings. This technology is already applied at staggering scales (\citet[][]{nguyen_colloidal_2004} estimated that a total of 2 billion tons of ore are treated annually). Given especially the relevance in the mining of copper, and in view of the strong push for electrification in response the the climate crisis, these numbers are likely to continue to rise in the future \citep{rogich_global_2008,world_bank_group_growing_2017}. The interest to understand the collision process better is driven by the demand for more reliable process modelling \citep{kostoglou_generalized_2020} but also by the need for performance improvement. The latter is especially a concern for small particles with diameters smaller than 20$\mu m$, where recovery is poor owing to their low collision rates \citep{nguyen_demonstration_2006,miettinen_limits_2010}.
	
	For modelling purposes, the collision process is generally separated into two components \citep{pumir_collisional_2016}: the `geometric collision rate', which considers the collisions neglecting any interaction between the collision partners,
	and the `collision efficiency', which quantifies how many of these collisions actually happen when taking the local modification of the flow field into account. The focus here is on the (ensemble-averaged) geometric collision rate between two species, `1' and `2', which when expressed per unit volume can be written as 
	\begin{equation}	\label{eq:CollisionKernelDynamic}
	Z_{12} = \Gamma_{12} n_1 n_2,
	\end{equation}
	where $\Gamma_{12}$ is the collision kernel, and $n_1$ and $n_2$ denote the respective number densities of the two species. The collision kernel measures the rate at which the separation vector between particle centres crosses the collision distance. For spherical particles with a collision distance $r_c = r_1+ r_2$ (where $r_1,r_2$ denote the particle radii), $\Gamma_{12}$ can be expressed as \citep{sundaram_collision_1997}
	\begin{equation}	\label{eq:CollisionKernelKinematic_4pi}
	\Gamma_{12} = 4\pi r_c^2g(r_c)S_-(r_c),
	\end{equation}
	where besides the surface area $4\pi r_c^2$ of the collision sphere, the other factors are the radial distribution function (RDF) at collision distance $g(r_c)$, which describes variations in the local particle concentration, and the effective radial approach velocity at contact,
	\begin{equation}    \label{eq:Smin}
	S_-(r_c) = -\int_{-\infty}^{0} \Delta v_r\mathrm{p.d.f.}(\Delta v_r|r_c) \mathrm{d}(\Delta v_r).
	\end{equation}
	Here, $\Delta v_r$ is the radial component of the relative velocity, which is positive when the pair separates, and $\mathrm{p.d.f.}(\Delta v_r|r_c)$ is the probability density function of $\Delta v_r$ conditioned on a pair with separation $r_c$. 
	
	The bidisperse collision kernel $\Gamma_{12}$ depends on a multitude of parameters characterising properties of the suspended particles and of the carrier flow. The discussion here is restricted to homogeneous isotropic turbulence for which the most relevant dependencies may be summarised in non-dimensional form as
	\begin{equation}
	\frac{\Gamma_{12}}{(r_c^3/\tau_\eta)} = f(St_1,St_2,\frac{\rho_1}{\rho_f},\frac{\rho_2}{\rho_f},\Rey_\lambda, \frac{r_1}{\eta},\frac{r_2}{\eta},Fr,\dots ). 
	\label{eq:Gammadep}
	\end{equation}
	Here, the Stokes number 
	\begin{equation}    \label{eq:StokesNumber}
	St_i = \frac{\tau_i}{\tau_\eta} = \frac{r_i^2(2\rho_i/\rho_f + 1)}{9\nu\tau_\eta},
	\end{equation}
	($i = 1,2$) characterises how well particles follow the flow by relating the particle response time $\tau_i = {r_i^2(2\rho_i/\rho_f + 1)}{/(9\nu)}$ to the Kolmogorov time scale of the turbulence $\tau_\eta = (\nu/\varepsilon)^{1/2}$, with $\nu$ and $\varepsilon$ denoting the kinematic viscosity and the average rate of turbulent dissipation, respectively. Further, the ratios of particle ($\rho_i$) and fluid ($\rho_f$) densities are relevant as they characterise to what extent the particle motion is influenced by fluid (`added mass') inertia and buoyancy. The particle radii $r_i$ determine the collision radius $r_c$, and their size relative to the Kolmogorov length scale $\eta = (\nu^3/\varepsilon)^{1/4}$ determines the range of turbulent scales relevant for their motion. In addition to turbulent driving, particle motion may also be affected by gravitational effects, and the relative importance of these two factors is captured by the Froude number $Fr = a_\eta/g$, where $a_\eta = \eta/\tau_\eta^2$ and $g$ are the turbulence and gravitational accelerations, respectively. Finally, the intensity of the turbulence is measured by the Taylor Reynolds number $\Rey_\lambda = \sqrt{15/(\nu \varepsilon)}u'^2$, where $u'$ is the single-component root-mean-square (r.m.s.) fluid velocity.
	Obviously, the entire parameter space spanned by (\ref{eq:Gammadep}) is vast and cannot be studied comprehensively here. We therefore limit the present investigation to cases with $St_1 = St_2 = St$, and to the zero-gravity regime, i.e. $Fr \to \infty$. The benefit of these choices lies in the fact that they keep the problem simple enough to disentangle the relevant mechanisms. Similarly, these configurations are the most amenable to modelling approaches and therefore allow for their evaluation at the most basic level. 
	
	Our investigation is based on direct numerical simulations of bubbles and particles in statistically stationary homogeneous isotropic turbulence using the point-particle approach. Details of this approach will be described in \S\ref{sec::methods} after first reviewing relevant modelling approaches for the collision kernel in \S\ref{sec::intro_bidispersedcol}.
	The results are shown in \S\ref{sec::results}, followed in \S\ref{sec::discussion&conclusion} by practical considerations in light of our results, and conclusions.
	
	\section{Theoretical background and existing models}
	\label{sec::intro_bidispersedcol}
	\subsection{The tracer limit $St \rightarrow 0$: shear mechanism}
	In the tracer limit of $St \to 0$, the suspended species follow the flow faithfully and distribute uniformly. This means that collisions occur only if particles of finite size are moved relative to each other due to shearing motions in the flow. Considering the dominant shear contribution of the smallest (Kolmogorov) scales of turbulence, and assuming local isotropy as well as Gaussian distribution of the flow velocity gradient, \citet{saffman_collision_1956} derived the classical result
	\begin{equation}	\label{eq:SaffmanZNoRDF}
	\Gamma_{12}^{(ST)} = \sqrt{\frac{8\pi}{15}}\frac{r_c^3}{\tau_\eta}
	\end{equation}
	predicting the rate of shear-driven collisions in a turbulent flow. Note that here we use the spherical formulation for the collision kernel, which was shown to be the appropriate form by \citet[][]{wang_reconciling_2005}. Employing the concept of a collision cylinder, instead, results in a slightly different value of the prefactor (\makered{$\sqrt{8\pi}/3\approx1.67$} instead of $\sqrt{8\pi/15} \approx 1.29$).
	
	\subsection{Intermediate $St$: preferential sampling and velocity decorrelation}\label{sec::intro_intermediateSt}
	For non-zero $St$, the suspended species no longer completely follow the flow. Such inertial effects influence the collision rate via two different pathways. First, even if the drift is small, its accumulated effect leads to preferential concentration, inducing clustering in the particle field. Additionally, the increasing decorrelation between the local particle and fluid velocities affects the collision velocities.
	
	Preferential concentration of inertial particles is widely observed experimentally \makered{\citep{aliseda_effect_2002,monchaux_preferential_2010,obligado_preferential_2014,petersen_experimental_2019,li_evidence_2021}} and numerically \citep{bec_heavy_2007,goto_self-similar_2006,ireland_effect_2016,voskuhle_prevalence_2014,calzavarini_dimensionality_2008}. It is rather straightforward to extend (\ref{eq:SaffmanZNoRDF}) to account for \makered{this effect} by simply multiplying it with the RDF $g_{12}(r_c)$ \citep{voskuhle_prevalence_2014}, yielding
	\begin{equation}	\label{eq:SaffmanZ}
	\Gamma_{12}^{(STc)} = \sqrt{\frac{8\pi}{15}}\frac{r_c^3}{\tau_\eta}g_{12}(r_c).
	\end{equation}
	Various approaches have been proposed to explain the phenomenon of preferential concentration \citep{chen_turbulent_2006,goto_self-similar_2006,coleman_unified_2009,bec_clustering_2005,bec_heavy_2007,fouxon_distribution_2012,maxey_gravitational_1987,zaichik_statistical_2009}. Among these, the most intuitive one is the `centrifuge picture' \citep{maxey_gravitational_1987}, according to which heavy particles are ejected out of eddies due to their inertia and hence accumulate in regions of low vorticity and high strain. For collisions between heavy particles (e.g. cloud droplets), it is therefore found that $g_{12}(r_c)\geq 1$ generally \citep{zhou_modelling_2001}, i.e. clustering enhances the collision rate in some cases even by multiple orders of magnitude \citep{ireland_effect_2016,voskuhle_prevalence_2014,pumir_collisional_2016}.
	
	For the collision velocity, inertial effects can be split into a local mechanism and a non-local mechanism. The local mechanism results from the fact that particles react differently to the same fluid forcing provided that they have different properties. Hence this effect contributes an additional relative velocity for bidispersed collisions only, and plays no role in monodispersed cases. For this to occur, the particle trajectories should not deviate significantly from the pathlines (i.e. $St$ not too large). An extension of the Saffman--Turner approach accounting for this effect has been reported by \citet{yuu_collision_1984}.
	
	In contrast, the non-local mechanism refers to the situation in which particles arrive at the same location with different particle velocities. Illustratively, one can think of these particles as being `slung out' of neighbouring eddies, and the effect is therefore also known as the `sling effect' \citep{falkovich_acceleration_2002}.
	Unlike the local mechanism, the sling effect is also active for monodisperse suspensions, and so far has been studied almost exclusively in this context \citep[e.g.][]{bewley_observation_2013,falkovich_acceleration_2002,falkovich_sling_2007,voskuhle_prevalence_2014,wilkinson_caustic_2006,ijzermans_segregation_2010}. From these studies, it has become clear \citep[see e.g.][for an overview]{pumir_collisional_2016} that the sling effect can significantly enhance monodisperse  collision rates by increasing $S_-$. A widely used parametrisation is $S_- \sim u_\eta F(St,\Rey_\lambda)$, where $u_\eta = \eta/\tau_\eta$ is the Kolmogorov velocity, such that the sling-induced collision rate is given by
	\begin{equation}
	\Gamma_{11}^{(slg)} = 4\pi r_c^2  u_\eta F(St,\Rey_\lambda).
	\label{eq:Gsling}
	\end{equation}
	It has then been proposed \citep{wilkinson_caustic_2006,voskuhle_prevalence_2014} to obtain the overall collision rate from the sum
	\begin{equation}
	\Gamma_{11}^{(tot)} = \Gamma_{11}^{(STc)} + \Gamma_{11}^{(slg)}.
	\label{eq:Gtot}
	\end{equation}
	The underlying idea for this decomposition is that particles that are clustered close to each other collide with low velocities, whereas those with high relative velocities can be assumed to be uniformly distributed. Note that both (\ref{eq:Gsling}) and (\ref{eq:Gtot}) are given for the monodisperse case only, since related results for the bidisperse case have not been reported yet.
	
	\subsection{Large $St$ limit: kinetic gas behaviour}
	At very large (but finite) $St$, the velocities of the suspended particles arriving at the same point are increasingly uncorrelated. Assuming entirely random and isotropic particle velocities in the spirit of the kinetic gas theory, \citet{abrahamson_collision_1975} derived the collision kernel
	\begin{equation}	\label{eq:Abrahamson12}
	\Gamma_{12}^{(A)} =  \sqrt{8\pi}r_c^2\sqrt{v_1'^2 + v_2'^2},
	\end{equation}
	where the mean-square single-component particle velocity $v_i'^2$ is related to flow properties via
	\begin{equation}	\label{eq:AbrahamsonVp}
	v_i'^2 = A_iu'^2 = \frac{\frac{T_{fL}}{\tau_i}+\gamma_{i}^2}{\frac{T_{fL}}{\tau_i}+1}u'^2,
	\end{equation}
	with $T_{fL}$ denoting the \makered{fluid} Lagrangian integral time scale, and $\gamma_i=3\rho_f/(2\rho_i + \rho_f)$.
	It has been pointed out \citep{voskuhle_prevalence_2014} that (\ref{eq:Abrahamson12}) is not strictly valid for turbulence as it fails to account for the multiscale structure of the flow. Alternative derivations \citep{volk_collisions_1980,mehlig_colliding_2007} based on the \citet{kolmogorov_local_1941} phenomenology arrived at \makered{$F(St,\infty) \sim K\sqrt{St}$}, with $K$ being a universal dimensionless constant, at the limit of intense turbulence in the context of (\ref{eq:Gsling}).
	
	\subsection{Modelling approaches for bubble--particle collisions in turbulence}\label{sec::intro_bpmodels}
	Next, we outline briefly the different approaches to modelling bubble--particle collisions reported in particular in the mining literature so far. For additional details, we refer the reader to the reviews on the topic \citep[e.g.][]{kostoglou_critical_2020,nguyen2016,hassanzadeh_review_2018}. Note that here and in the following we use subscripts $i = b$,$p$ to denote bubbles and heavy particles, respectively. 
	
	The most commonly adopted approach is to assume the high-$St$ limit and to base the collision models on (\ref{eq:Abrahamson12}). The differences from the theory of \citet{abrahamson_collision_1975} are related to the expressions used to model the r.m.s. velocities. Instead of (\ref{eq:AbrahamsonVp}), \citet{schubert_turbulence-controlled_1999} and later \citet{bloom_structure_2002} used the relation given by \citet{liepe_untersuchungen_1976}, which reads
	\begin{equation}    \label{eq:LiepeMoekel}
	w_i'^{(LM)} = 0.57\frac{\varepsilon^{4/9}r_i^{7/9}}{\nu^{1/3}}\bigg(\frac{\rho_f-\rho_i}{\rho_f}\bigg)^{2/3}.
	\end{equation}
	This result is based on an analogy to gravitational settling with the fluid acceleration in the inertial range replacing gravity. The resulting apparent weight is balanced by Allen's drag, which scales with the particle slip velocity $\boldsymbol{w_i} = \boldsymbol{v_i} - \boldsymbol{u}$ as $|\boldsymbol{w_i}|^{3/2}$ (with boldface denoting vectors). Here, $\boldsymbol{v_i}$ is the bubble/particle velocity, and $\boldsymbol{u}$ is the flow velocity.
	Later work by \citet{nguyen_colloidal_2004} used (\ref{eq:LiepeMoekel}) with a different constant, 0.83, for bubbles. For particles, they replaced the inertial subrange acceleration with that in the dissipation range, and Allen's drag with Stokes drag, in order to account for the small size of typical particles. This resulted in
	\begin{equation}    \label{eq:NS}
	w_p'^{(NS)} = \frac{2r_p^3\varepsilon}{135\nu^2}\bigg(\frac{\rho_p-\rho_f}{\rho_f}\bigg).
	\end{equation}
	Importantly, both (\ref{eq:LiepeMoekel}) and (\ref{eq:NS}) are expressions for the relative velocity between bubbles/particles and the surrounding fluid. Their use with Abrahamson's theory is therefore inconsistent since (\ref{eq:Abrahamson12}) contains velocities in a fixed frame of reference. This was already pointed out in \citet{kostoglou_generalized_2020}. We further note that the quasi-static assumption underlying (\ref{eq:LiepeMoekel}) and (\ref{eq:NS}) is equivalent to a low-$St$ approximation and therefore not valid for the high-$St$ limit in which (\ref{eq:Abrahamson12}) applies. In fact, (\ref{eq:NS}) is consistent with the rigorously derived small-$St$ limit \citep{fouxon_distribution_2012}
	\begin{equation}	\label{eq:FouxonSlip}
	\boldsymbol{w_i}=\beta_i\tau_i \boldsymbol{a_f},
	\end{equation}
	with
	\begin{equation}
	\beta_i = \frac{2(\rho_f - \rho_i)}{\rho_f + 2\rho_i},
	\end{equation}
	if the fluid acceleration is approximated by the dissipative scaling $|\boldsymbol{a}_f| \approx \varepsilon r_i/(15 \nu)$. Note, however, that it appears more appropriate to use $|\boldsymbol{a}_f| \approx a_\eta$ for small particles ($r_i/\eta \lessapprox 1$) because $|\boldsymbol{a}_f| \rightarrow 0$ otherwise. \makeredTwo{In the general case, the fact that the fluid acceleration experienced by the particle may vary considerably over the particle response time precludes a simple relation between $\boldsymbol{w_i}$ and $\boldsymbol{a_f}$.}
	
	In a different approach, \citet{ngo-cong_isotropic_2018} extended the model by \citet{yuu_collision_1984} to the bubble--particle case. The resulting expression takes the form
	\begin{equation}	\label{eq:NgoCongModel}
	\Gamma_{bp}^{(NC)} = (r_b + r_p)^2\sqrt{\frac{8\pi}{3}\bigg\{3(A_b + A_p - 2B)u'^2 + (A_br_b^2 + A_pr_p^2 + 2Br_pr_b)\frac{\varepsilon}{3\nu}\bigg\}},	
	\end{equation}
	where the term proportional to $u'^2$ represents the `local' inertial effect on the relative velocity that adds to the shear-driven collisions that are accounted for by the $\varepsilon$ term. Aside from bubble/particle properties, the coefficients $A_i$ and $B$ depend also on $T_{fL}$, similar to (\ref{eq:AbrahamsonVp}). The bubble--particle velocity correlation is determined by $B$. \citet{ngo-cong_isotropic_2018} additionally incorporated (\ref{eq:LiepeMoekel}) and (\ref{eq:NS}) into this model. However, doing so suffers from the same inconsistencies outlined above. As a consequence, this expression featured a negative radicand when evaluated for the parameters in this study, and is therefore not included further. \makered{It is worth mentioning that the term proportional to $u'^2$ in (\ref{eq:NgoCongModel}) does not approach (\ref{eq:Abrahamson12}) even at large $St$, since $A_i=B$ for identical particles. We have therefore extended the model along the lines of the theory of \citet{kruis_collision_1997}, which attempted to reconcile this issue by marrying the concept of \citet{yuu_collision_1984} for the small $St$ case and that of \citet{williams_particle_1983} (which does not consider shear-driven collisions) for the large $St$ case, to bubble--particle collisions (see Appendix \ref{sec::extKruisKusters}). The original formulation of \citet{kruis_collision_1997} accounted for collisions of particles of equal density only.}
	We further note the work of \citet{fayed_direct_2013}, who employed the model of \citet{zaichik_turbulent_2010}, which was developed originally for collisions between particles with arbitrary but equal density assuming a joint-normal fluid--particle velocity distribution, to the bubble--particle case. 
	Recently, \citet{kostoglou_generalized_2020} proposed another model that is specific to the case of very fine particles that essentially follow the flow, such that the relative velocity is dominated by the bubble slip velocity.
	
	A common feature of almost all these models is that they are direct adaptations of concepts developed for collisions between heavy particles. They therefore fail to account for fundamental differences in \makered{how bubbles and heavy particles react to a turbulent flow. Most strikingly, for example,} the response to fluid accelerations given in (\ref{eq:FouxonSlip}) is in opposite directions as $\beta$ switches sign from negative to positive from $\rho_i>\rho_f$ to $\rho_i<\rho_f$. Similarly, applying the centrifuge picture to light particles such as bubbles, one expects them to concentrate in regions of high vorticity as they travel to the centre of the eddies. This implies that heavy and light particles segregate in a turbulent flow, as has indeed been observed \citep{calzavarini_quantifying_2008-1,fayed_direct_2013}. As a consequence, it is expected that $g_{bp}(r_c)< 1$ in these cases, such that preferential concentration is expected to lead to a decrease in $\Gamma_{bp}^{(STc)}$ for collisions between heavy and light particles. Such aspects cannot be quantified easily from laboratory-scale flotation set-ups \citep[e.g.][]{darabi_investigation_2019} since it is difficult to disentangle the many factors influencing the overall flotation rate. Numerically, \citet{wan_study_2020} observed no reduction of the RDF, but increased relative velocities for bubble--particle pairs based on a point-particle simulation while taking gravitational effects into account.
	For cases without gravity, \citet[][]{fayed_direct_2013} reported $g_{bp}(r_c) < 1$ and relative velocities matching those predicted by the theory of \citet{zaichik_turbulent_2010} for simulations of the extreme case $\rho_b/\rho_f = 0 $ and $\rho_p/\rho_f\to \infty $ across a range of $\tau_i$. It is the goal of the present study to add to this a systematic investigation of how bubble--particle relative behaviour affects their collision statistics in the low- to moderate-$St$ regime, and to assess how this affects modelling outcomes.
	
	\section{Methods}\label{sec::methods}
	\subsection{Fluid phase} \label{subsec:fluid_phase}
	To obtain the background turbulence, we solve the Navier--Stokes equation and the continuity equation for incompressible flow:
	\begin{subeqnarray}
		\frac{\mathrm{D}\boldsymbol{u}}{\mathrm{D}t} &=& -\frac{1}{\rho_f}\nabla P + \nu \nabla^2\boldsymbol{u} + \boldsymbol{f_\Psi},\\[3pt]
		\nabla\boldsymbol{\cdot}\boldsymbol{u} &=& 0,
	\end{subeqnarray}
	where $\mathrm{D}/\mathrm{D}t$ is the material derivative following a fluid element, $t$ is the time, and $P$ is the pressure. The forcing $\boldsymbol{f_\Psi}$, which is non-zero only for the wavenumbers $|\boldsymbol{\kappa}|/|\boldsymbol{\kappa_0}| < 2.3$ (i.e. the largest scales), with $|\boldsymbol{\kappa_0}|$ being the smallest wavenumber along each direction, is added to counter dissipation and maintain statistical stationarity. We employ the widely used \citet{eswaran_examination_1988} forcing scheme \citep{chouippe_forcing_2015,spandan_fluctuation-induced_2020}. In brief, a complex vector is generated in Fourier space for the forced wavenumbers by multiple Uhlenbeck--Ornstein processes. This vector is then projected onto the plane normal to the wavevector, thereby ensuring that $\boldsymbol{f_\Psi}$ is divergence-free. To simulate the fluid motion, a second-order finite-difference solver is implemented on a staggered grid \citep{verzicco_finite-difference_1996,van_der_poel_pencil_2015}. All spatial derivatives, including the nonlinear terms, are discretised by a second-order central finite-difference method. The simulation domain is a cubic box with length $L_{box}=1$ and triply periodic boundary conditions to eliminate boundary effects. Time marching is performed using a fractional step third-order low-storage Runge--Kutta scheme and the implicit Crank--Nicolson scheme for all viscous terms at a maximum Courant--Friedrichs--Lewy (CFL) number 1.2. The CFL number is $\max(|u_1| + |u_2| + |u_3|)\Delta t/\Delta x$ over all cells, where the grid spacing $\Delta x$ is identical in all the three dimensions. Here, $u_1,u_2,u_3$ are the \mbox{$x$-,} \mbox{$y$-,} $z$-components of $\boldsymbol{u}$, and $\Delta t$ is the time step. The simulation is parallelised via slab decomposition along the $z$-direction.
	
\begin{figure}
		\centerline{\includegraphics{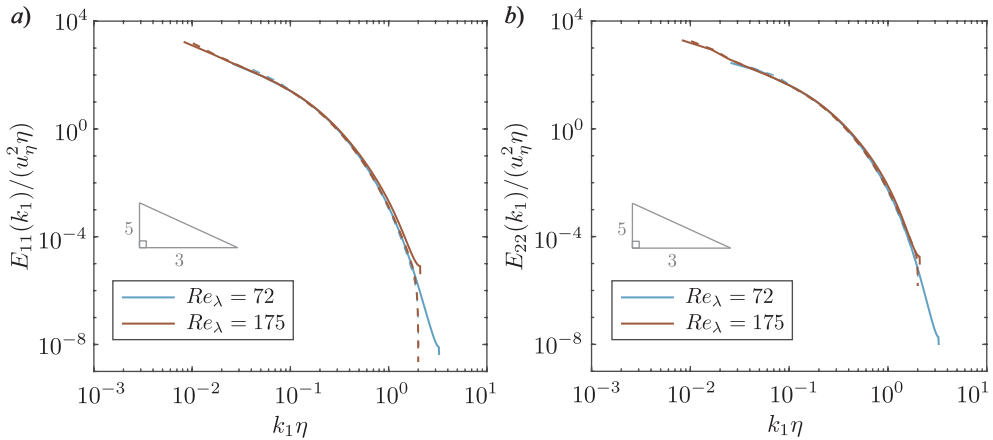}}
		\caption{The (\textit{a}) longitudinal and (\textit{b}) transverse energy spectra in single-phase statistically stationary homogeneous isotropic turbulence. Dashed lines show the data from \citet{jimenez_structure_1993}; triangles represent the $-5/3$ power law. The agreement with the literature is excellent, so the dashed lines can be obscured, especially in (\textit{b}).}
		\label{fig:Pspec}
	\end{figure}

	Turbulence with $\Rey_\lambda = 72$ and $175$ is generated from an initially quiescent fluid following the tuning method proposed by \cite{chouippe_forcing_2015}. The simulations are allowed to run until the pseudo-dissipation $\overline{\varepsilon}$ and $\Rey_\lambda$ are statistically stationary over at least $30T_L$. These statistically stationary flow fields are then used as the starting fields for the point-particle simulations. The flow statistics are listed in table \ref{tab:singlePhaseStat}. For validation, the longitudinal and transverse energy spectra are plotted in figure \ref{fig:Pspec}. Excellent agreement with the literature is shown.
	
	\begin{table}
		\begin{center}
			\begin{tabular}{lccccccccc}
				$\Rey_\lambda$  & $\mathcal{N}$   &   $\overline{\varepsilon}$ & $\eta$ & $k_{max}\eta$ & $u_\eta$ & $u'/u_\eta$ & $u_x'/u_y'$ & $T_L/\tau_\eta$ & $N_{b,p}$\\[3pt]
				72 & $256^3$ & 58.0 & 0.0041 & 3.3 & 0.62 & 4.3 & 0.98 & 19 & 10000\\
				175 & $512^3$ & 349 & 0.0013 & 2.1 & 0.77 & 6.6 & 0.97 & 43 & $77700-140000$\\
			\end{tabular}
			\caption{Statistics of the homogeneous isotropic turbulence: the grid size ($\mathcal{N}$), pseudo-dissipation ($\overline{\varepsilon}$), Kolmogorov length scale ($\eta$), maximum wavenumber ($k_{max}$), Kolmogorov velocity ($u_\eta$) scale, root-mean-square velocity fluctuations ($u'$), large-scale isotropy ($u_x'/u_y'$), and the large eddy turnover time ($T_L$) \makered{relative to the Kolmogorov time scale ($\tau_\eta$)}. $N_{b,p}$ are the numbers of bubbles and particles respectively.}
			\label{tab:singlePhaseStat}
		\end{center}
	\end{table}
	
	\subsection{Suspended phases} \label{subsec:suspended_phase}
	Bubbles and particles in the system are modelled using the point-particle approximation, where forces act on point masses. The bubble and particle dynamics are governed by \citep{maxey_equation_1983,tchen_mean_1947}
	\begin{equation}\label{eq:SimpleParticleEoM}
	\frac{4}{3}\pi r_i^3\rho_i\frac{d\boldsymbol{v_i}}{dt} = 6\pi\mu r_if_i(\boldsymbol{u}-\boldsymbol{v_i}) + \frac{4}{3}\pi r_i^3\rho_f\frac{\mathrm{D}\boldsymbol{u}}{\mathrm{D}t} + \frac{2}{3}\pi r_i^3\rho_f\bigg(\frac{\mathrm{D}\boldsymbol{u}}{\mathrm{D}t} - \frac{d\boldsymbol{v_i}}{dt}\bigg),
	\end{equation}
	where $\mu=\nu\rho_f$ is the absolute viscosity\makered{, $r_i$ is determined from $St_i$, and $\boldsymbol{u}$ is evaluated at particle position in this equation}.
	The three terms on the right-hand side of (\ref{eq:SimpleParticleEoM}) are the drag force, the pressure gradient force and the added mass force, respectively. \makered{Note that history forces, lift and reverse coupling are not considered to render it easier to disentangle the relative behaviour of bubble and particles. Nevertheless, as the lift force can be expected to play a role for bubbles, its effect will be discussed in \S\ref{sec::LiftDensityDrag}.} \makered{Furthermore, a one-way coupled system is a consistent choice to study the geometric collision rate where hydrodynamic interactions are neglected. It entails the assumption of the dilute limit in which turbulence modifications by the suspended species are negligible \citep{brandt_particle-laden_2022}.}
	The correction factor $f_i = 1 + 0.169Re_i^{2/3}$ accounts for finite bubble or particle Reynolds number $Re_i = 2r_i|\boldsymbol{w_i}|/\nu$, \makered{and implies the assumption of rigid spheres that obey the no-slip boundary condition for both species \citep{nguyen_colloidal_2004}. This is realistic since liquids in flotation cells typically contain significant amounts of surfactants, so that the bubble surfaces would likely be contaminated \citep{nguyen_colloidal_2004,huang_effect_2012}. Although other commonly used expressions for $f_i$ are available, the difference is minimal, as shown in \makered{Appendix \ref{sec::DragCorrection}}.}
	
	Equation (\ref{eq:SimpleParticleEoM}) is solved for each bubble and particle using a finite-difference scheme. To determine the flow velocities and the velocity gradients at bubble and particle positions, \makered{these quantities are interpolated from the Eulerian grid of the fluid solver to the particle positions using tri-cubic Hermite spline interpolation with a stencil of four points per direction.} This choice is made as Hermite splines are comparable in accuracy \citep{van_hinsberg_enhanced_2017} and computationally cheaper to implement than B-splines \citep{ostilla-monico_multiple-resolution_2015}. Time marching of $\boldsymbol{v_i}$ is performed with the explicit forward Euler method, and that of the positions of the suspended phases is done using the second-order Adams--Bashforth scheme. For stability, the time \makered{step is restricted such that neither the fluid nor the particle CFL number ($\max(v_{i1},v_{i2},v_{i3})\Delta t/\Delta x$) exceeds the value 1.2. This limit is enforced} in both the suspended- and fluid-phase solvers, which run with the same simulation time step. \makered{We have compared particle statistics to data from the literature in order to verify our code, and those results are included in Appendix \ref{sec::AppCodeVerification}.}
	
	Collisions between particles and bubbles are treated as `ghost collisions'. Under this approach, a `collision' occurs once the centres of the members of an approaching pair reach the collision distance and the colliding pair pass each other without interaction. The collision radii are determined through the virtual radii $r_i$ as computed from $St_i$. This scheme is often employed by simulations of particles in turbulence \citep{ireland_effect_2016,voskuhle_prevalence_2014,bec_clustering_2005,bec_heavy_2007,goto_sweep-stick_2008} and has been shown to be consistent with the formulation of $\Gamma_{pp}^{(ST)}$ \citep{wang_collision_1998}. To suppress the effect of different length scales when comparing collision pairs, we take $r_c = r_b + r_p$ for every type of collision. \makered{This is unlike studies examining solely particle--particle collisions, which define $r_c = 2r_p$ \citep[e.g.][]{ireland_effect_2016,voskuhle_prevalence_2014}.} Numerically, these collisions are detected with the `proactive' detection scheme in \citet{sundaram_numerical_1996}.
	
	The details of the simulation of the suspended phases are as follows. \makered{Particles ($\rho_p/ \rho_f  = 5$) and bubbles ($\rho_b/ \rho_f  = 1/1000$)  with $St_i = 0.1-3$} were seeded randomly and homogeneously \makered{($10000-140000$ each)} into the turbulent flow after it had reached a statistically stationary state as described in \S\,\ref{subsec:fluid_phase}. \makered{The density ratios correspond to sulphide minerals colliding with air bubbles in water, while the simulated $St$ range falls within $0.1 \lesssim St \lesssim 100$, which corresponds to particle sizes yielding the highest mineral recovery rate in conventional flotation cells ($10\mu$m$\lesssim r_p\lesssim 100\mu$m when $\nu = 10^{-6}$m\textsuperscript{2}/s and $\varepsilon = 5$m\textsuperscript{2}/s\textsuperscript{3}, following \citet{ngo-cong_isotropic_2018}).} For each case, bubbles and particles have identical $St_i$ to \makered{focus on the effect of different densities and to keep the parameter space manageable}. The upper limit $St_i \leq 3$ is mandated by the fact that the increasingly large (virtual) bubble radius for larger $St$ violates the point-particle approximation. For $St_b = 3$ and at $\Rey_\lambda = 175$, we have $r_b/\eta \approx 5$, which is already marginal \citep{homann_finite-size_2010}. We monitor the p.d.f.s of bubble and particle positions for statistical stationarity. Once this has been reached, collision statistics are collected over at least $7.7T_L$. Two kinds of statistics are evaluated: in a fixed reference frame considering all particles/bubbles (Eulerian statistics), and along individual trajectories of the pairs that collide (Lagrangian statistics). The former are computed at least every $\sim0.06T_L$, while the latter are calculated every time step (with only 25\% of all bubbles for the bubble--bubble case as the number of colliding bubbles is high). All the statistics presented are time- and ensemble-averaged unless indicated otherwise.
	
	\section{Results}\label{sec::results}
	\subsection{Eulerian statistics}\label{sec::EulerianStat}
	\subsubsection{Collision kernel}
	\begin{figure}
		\centerline{\includegraphics{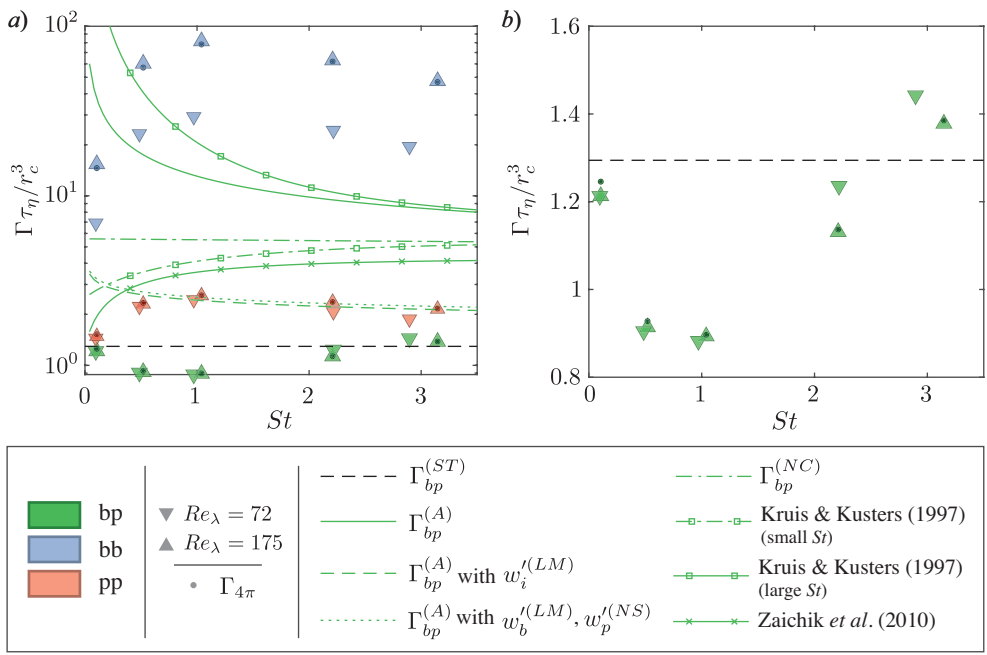}}
		\caption{(\textit{a}) The dimensionless bubble--particle (bp), bubble--bubble (bb) and particle--particle (pp) collision kernels at collision distance. The triangle symbols denote $\Gamma$ determined directly from the collision rate, while $\Gamma_{4\pi}$ (shown only for the $\Rey_\lambda = 175$ cases) is calculated according to (\ref{eq:CollisionKernelKinematic_4pi}). Unless specified otherwise, the colour conventions for all figures follow this figure. (\textit{b}) Zoomed-in version of (\textit{a}) plotted on linear scale.}
		\label{fig:collK}
	\end{figure}
	
	Figure \ref{fig:collK}(\textit{a}) shows simulation results for the dimensionless bubble--particle (bp), bubble--bubble (bb) and particle--particle (pp) collision kernels $\Gamma$. In addition to determining $\Gamma$ based on counting the number of collisions per time step (solid triangles), the collision kernels for $\Rey_\lambda = 175$ are determined indirectly via the RDF and the effective radial collision velocity according to (\ref{eq:CollisionKernelKinematic_4pi}) (shown as dots). Both results match closely, \makered{verifying} our analysis procedure. With the normalisation by $\tau_\eta/r_c^3$ suggested by the Saffman--Turner framework, the results are insensitive to the change in $\Rey_\lambda$ for the pp and bp cases, but not for bb collisions. 
	From the data, it is further evident that the relative behaviour of bubbles and particles is distinct from that of identical particles. For collisions between identical species, $\Gamma\tau_\eta/r_{c}^3$ is maximum when $St \sim 1$, while $\Gamma_{bp}\tau_\eta/r_{c}^3$ exhibits a minimum for this value of $St$. This trend is not captured by any of the models for the bp case discussed in \S \ref{sec::intro_bpmodels}, for which the predictions are included as lines in figure \ref{fig:collK}(\textit{a}). Generally, the model predictions are also significantly higher than the actual collision rates obtained from the simulations, the exception being the \citet{saffman_collision_1956} model, which best captures the magnitude yet fails to predict the proper $St$-trend, as is illustrated more clearly in figure \ref{fig:collK}(\textit{b}).
	
\begin{figure}
		\centerline{\includegraphics{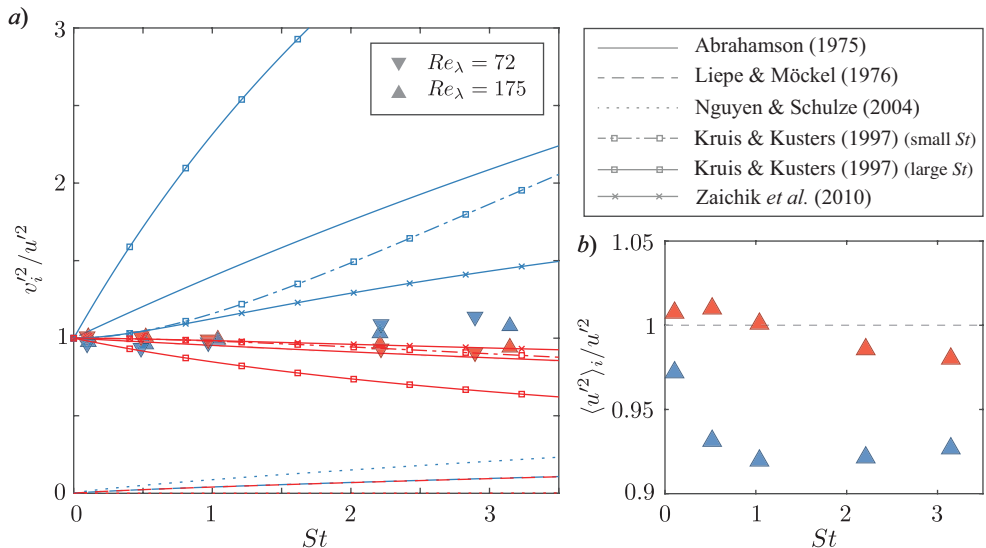}}
		\caption{\makered{(\textit{a})} The mean-square velocity of bubbles and particles at various $St$. Models are shown for $\Rey_\lambda = 175$ only. \makered{(\textit{b}) The mean-square fluid velocity conditioned at bubble/particle positions for $\Rey_\lambda = 175$.}}
		\label{fig:vrms}
	\end{figure}

	The fact that the \citet{abrahamson_collision_1975} \makered{and large-$St$ \citet{kruis_collision_1997} predictions do not match the data is to be expected as the present $St$ range does not match the assumptions made in these frameworks.} Naturally, this also transfers to all approaches based on $\Gamma_{bp}^{(A)}$, and the somewhat better agreement with our data for models employing (\ref{eq:LiepeMoekel}) and (\ref{eq:NS}) instead of (\ref{eq:AbrahamsonVp}) is rather an artefact of the inconsistencies discussed earlier. This is emphasised by figure \ref{fig:vrms}\makered{(\textit{a})}, where the large difference between the modelled slip velocities and the mean-square bubble/particle velocities $v_i'^2$ is obvious. Notably from the same figure, $v_p'^2$ is well predicted by the models \makered{of \citet{abrahamson_collision_1975}, \citet{kruis_collision_1997} (small $St$) and \citet{zaichik_turbulent_2010}.} \makered{On the other hand, model estimates for $v_b'^2$ are generally too high. It is insightful to note the stark overprediction by (\ref{eq:AbrahamsonVp}) as the underlying framework by \citet{abrahamson_collision_1975} is also employed in the models by \citet{ngo-cong_isotropic_2018}. The resulting overprediction of $v_b'^2$ might therefore explain in part why the estimate of $\Gamma_{bp}^{(NC)}$ is too large even though their concept (which is based on \citet{yuu_collision_1984}) is in principle more suitable for the moderate values of $St$ here. 
		Unlike \citet{abrahamson_collision_1975}, \citet{kruis_collision_1997} (small $St$) include dissipation range scaling in their modelling, which is seen to improve the prediction at $St\lessapprox 1$ but still overestimates $v_b'^2$ for larger $St$.}
	\makered{Finally, there is another factor, which affects all models. Bubbles especially do not sample the flow randomly, such that the mean-square fluid velocity at bubble locations $\langle u'^2\rangle_b$ is almost 10\% lower than $u'^2$, as shown in figure \ref{fig:vrms}(\textit{b}). This implies that bubbles sample flow regions with weaker fluid velocity fluctuations such that using $u'^2$ as model input may overestimate $ v'^2_b$. Figure \ref{fig:vrms}(\textit{b}) also shows that this effect is less pronounced for heavy particles. The underlying cause for these observations is preferential sampling of flow regions, which we will discuss in \S \ref{sec::spatialDistribution}.}
	
	\subsubsection{Bubble/particle spatial distribution}    \label{sec::spatialDistribution}
	\begin{figure}
		\centerline{\includegraphics[]{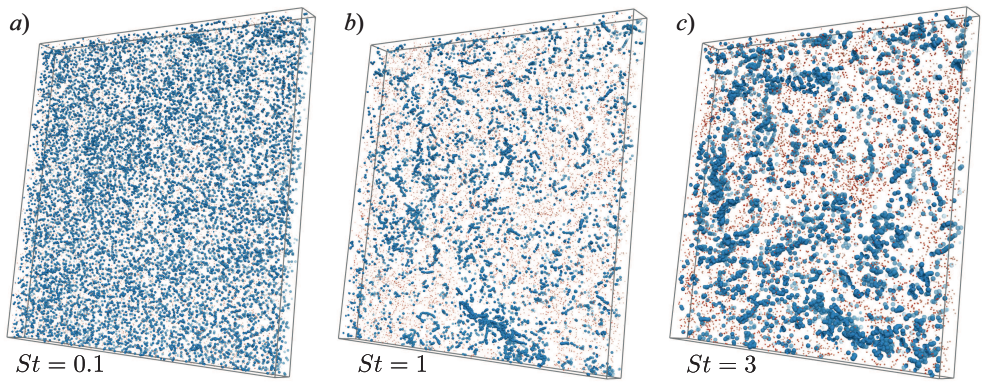}}
		\caption{Instantaneous snapshots of bubbles and particles in a slice with width $\times$ height $\times$ depth =  $L_{box} \times L_{box} \times 20\eta$ in the non-transient state for $\Rey_\lambda = 175$ at (\textit{a}) $St = 0.1$, (\textit{b}) $St = 1$, and (\textit{c}) $St = 3$. The size of the $St = 0.1$ bubbles and particles is tripled for visibility.}
		\label{fig:snapshot}
	\end{figure}
	
\begin{figure}
		\centerline{\includegraphics{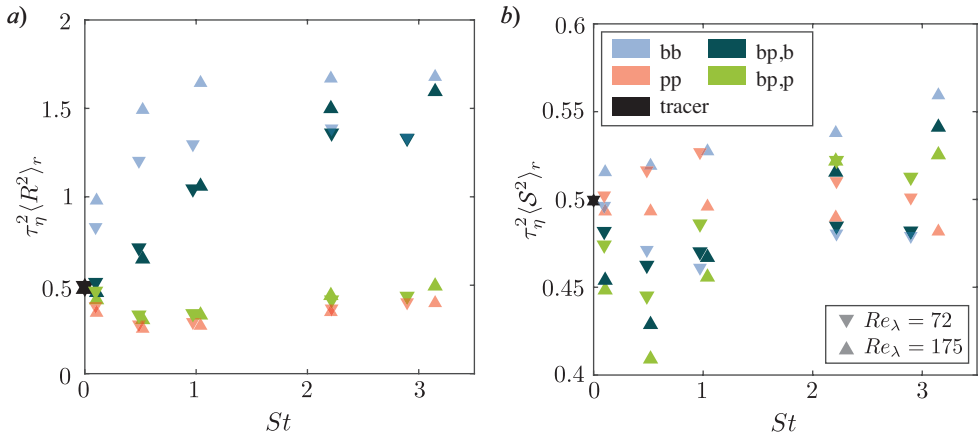}}
		\caption{Average value of the norm of (\textit{a}) the rotation rate $R^2$, and (\textit{b}) the strain rate $\mathcal{S}^2$, of the flow at bubble/particle/tracer positions. The bubble and particle data are conditioned on pairs with separation $r \leq 2r_{c}$. Here, bp,b(bp,p) refer to bubbles(particles) in bubble--particle pairs.}
		\label{fig:RsqSsq}
	\end{figure}

	In order to elucidate in particular the $St$ trends for $\Gamma_{bp}$, we first investigate the distribution of bubbles and particles in the flow. While all the models assume homogeneous distributions, this does not hold at intermediate $St$, as figure \ref{fig:snapshot} confirms, where instantaneous snapshots of the bubble and particle fields at different $St$ and $\Rey_\lambda=175$ are shown. For $St = 1$ and 3, bubbles and particles are seen to cluster but do so in different regions of the flow. This behaviour has been observed previously in the literature \citep{calzavarini_quantifying_2008-1,fayed_direct_2013,wan_study_2020} \makered{and is additionally shown by the different mean-square fluid velocity at bubble/particle positions in figure \ref{fig:vrms}(\textit{b})}. To investigate this preferential concentration, we plot the norm of the rotation $\langle R^2 \rangle_r$ and strain $\langle \mathcal{S}^2 \rangle_r$ rates of the flow at the particle/bubble positions in figure \ref{fig:RsqSsq}, with
	$\langle \cdot \rangle_r$ denoting an ensemble average over particles in pairs with separations smaller than $r$. For tracers ($St=0$), we obtain $\tau_\eta^2\langle R^2 \rangle_r=\tau_\eta^2\langle \mathcal{S}^2 \rangle_r =0.5$ with $r\to \infty$, consistent with the analytical result in statistically stationary homogeneous isotropic turbulence. For the other cases, $\langle R^2\rangle_r$ and $\langle \mathcal{S}^2\rangle_r$ are conditioned on pairs with $r\leq 2r_c$.  Figure \ref{fig:RsqSsq}(\textit{a}) shows that bubbles(particles) cluster in regions of high(low) rotation rate, which is consistent with the centrifuge picture. Due to the clustering, conditioning has little effect for monodisperse collisions, and the results are therefore very close to single particle statistics. This is different for bp pairs, where $R^2$ is consistently lower (higher) for bubbles (particles) close to a particle (bubble). These observations imply that bp collisions occur for a subset of bubbles/particles that is located outside of their respective `preferred' location within the flow.
	We note that correlations between particle/bubble locations and the strain rate (figure \ref{fig:RsqSsq}(\textit{b})) are much weaker and do not display the same qualitative trends observed for $R^2$. This is consistent with observations by e.g. \citet{ireland_effect_2016} and \citet{wang_reynolds_2020}, who also found only a weak correlation between the positions of heavy particles and regions of the flow with low strain rate.
	
\begin{figure}
		\centerline{\includegraphics{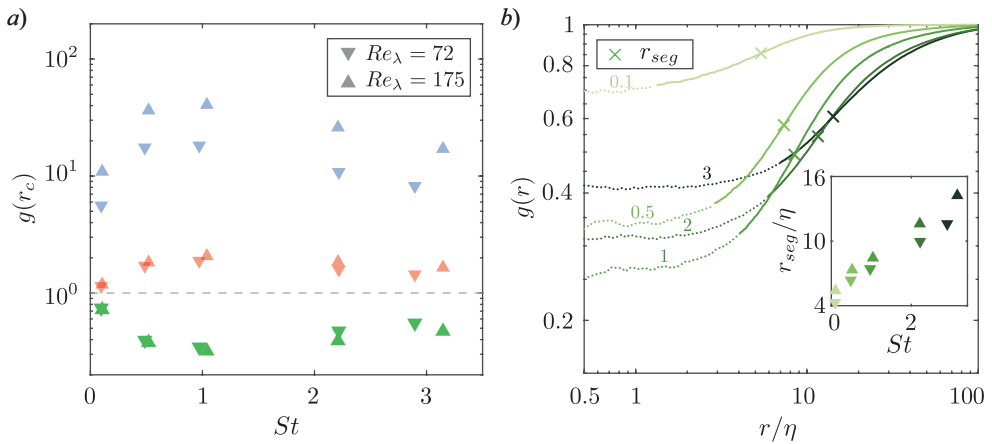}}
		\caption{(\textit{a}) RDF at collision distance. (\textit{b}) The bubble--particle RDF as a function of $r$ and the segregation length scale $r_{seg}$ \makeredTwo{at $\Rey_\lambda = 175$}. The number above each line is the corresponding $St$, and the dotted segments denote the RDF below the collision distance. Inset shows $r_{seg}$ at various $St$.}
		\label{fig:RDF}
	\end{figure}

	A consequence of the preferential concentration in different flow regions is that bubbles and particles become segregated. This effect is quantified by the RDF $g(r)$. Essentially, $g(r)$ relates the actual number of pairs with separation $r$ to that expected for uniformly distributed particles. Therefore, $g(r) = 1$ when particles are uniformly distributed, $g(r)>1$ implies clustering, and $g(r)<1$ implies segregation \citep{saw_studies_2008}. Figure \ref{fig:RDF}(\textit{a}) shows results for the RDF at collision distance from our simulations. Bubble--particle pairs indeed segregate, and the segregation is strongest when $St = 1$, as reflected by the local minimum of $g(r_{c})$. This is compatible with the corresponding behaviour of bubbles and particles, which always cluster for the tested parameters, with maximum clustering when $St\sim 1$. Increasing $\Rey_\lambda$ increases segregation slightly, but the effect is weak. 
	
	Figure \ref{fig:RDF}(\textit{b}) displays the bubble--particle RDF as a function of $r$ at $\Rey_\lambda=175$, which help us to understand the effect of different length scales. The shape of the $g(r)$ curves is distinctly different from the power-law behaviour reported for the monodisperse case \citep{ireland_effect_2016}.  For all $St$, there is an essentially flat region for $g(r)$ at small scales, followed by a transition region with the steepest gradient at intermediate scale $\sim 10\eta$ before approaching 1 for large separations. This behaviour suggests that the relevant length scale for the segregation, i.e. a typical distance between bubble and particle clusters, is at intermediate scales. 
	To quantify this more precisely, we define the separation corresponding to the point of inflexion in $g_{bp}(r)$ as the segregation length scale $r_{seg}$. The values of $r_{seg}$ are marked by crosses in figure \ref{fig:RDF}(\textit{b}), and plotted against $St$ in the inset. The figure shows that $r_{seg}/\eta$ increases approximately linearly with $St$ and increases slightly with $\Rey_\lambda$. These findings are consistent with those by \citet{calzavarini_quantifying_2008-1}, who studied segregation based on a concept inspired by Kolmogorov's distance measure \citep{kolmogorov_approximation_1963}. In particular, these authors also report segregation scales $\sim 10\eta$ with an increasing trend for higher $\Rey_\lambda$.
	
\begin{figure}
		\centerline{\includegraphics{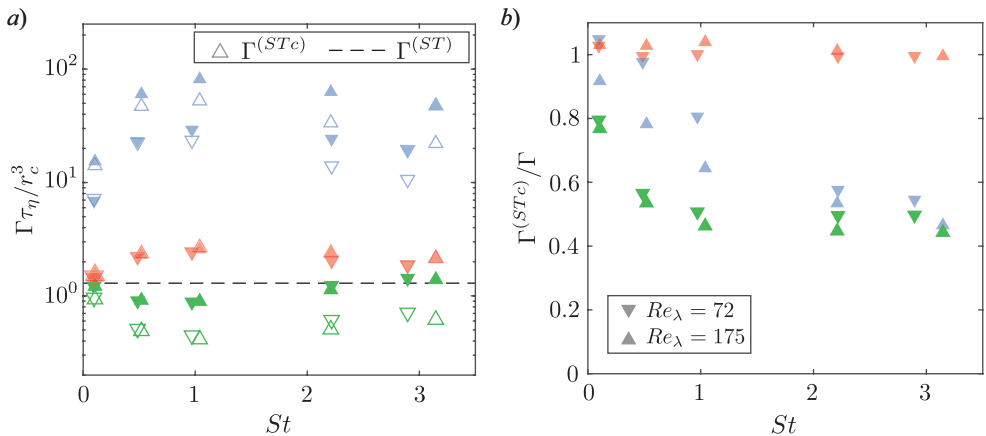}}
		\caption{(\textit{a}) The dimensionless collision kernel (solid symbols) and the Saffman--Turner prediction after accounting for the RDF $\Gamma^{(STc)}$ (hollow symbols). (\textit{b}) The ratio between $\Gamma^{(STc)}$ and the collision kernel.}
		\label{fig:collKShearOverCollK}
	\end{figure}

	The trends observed for $g(r_c)$ in figure \ref{fig:RDF}(\textit{a}) remarkably resemble those discussed for the $St$ dependence of $\Gamma$ earlier, in the context of figure \ref{fig:collK}. It is therefore suggestive to think that the effects of segregation may explain in particular the discrepancy between the data and $\Gamma^{(ST)}$. We can check this by plotting $\Gamma^{(STc)}$ as defined in (\ref{eq:SaffmanZ}), which is shown as hollow symbols in figure \ref{fig:collKShearOverCollK}(\textit{a}). From this plot, it can be seen that correcting $\Gamma^{(ST)}$ with the RDF leads to an almost perfect match with the pp collision kernel. \makered{This differs from, but is only seemingly at odds with, the results in \citet{voskuhle_prevalence_2014} due to the larger collision distance considered here as a consequence of keeping $r_c= r_b + r_p$ constant for all collision types. This therefore implies that while pp collisions at moderate $St$ are governed by the sling mechanism at small separations, they remain shear-dominated at larger ones.} In contrast, $\Gamma^{(STc)}$ overcorrects the bp collision kernel and undercorrects that of bb collisions. In figure \ref{fig:collKShearOverCollK}(\textit{b}), we replot the same data in the form of the ratio $\Gamma^{(STc)}/\Gamma$, which can be interpreted as the relative contribution of the shear mechanism (compensated for segregation/clustering) to the overall collision rate. For pp collisions, this ratio is very close to 1 throughout. The situation is different for the bb case, where $\Gamma^{(STc)}/\Gamma$ approaches 1 only for the lowest $St$ considered, and the value drops significantly for the higher $St$. Consistently the lowest values for $\Gamma^{(STc)}/\Gamma$ are observed for bp collisions where the value quickly drops to around 0.5, which implies that a significant part of the relative velocities cannot be explained by the shear mechanism in this case. It is important to stress here that keeping $r_c$ constant for all collision types for a given $St$ removes this as a factor, such that the observed trends across species can only be rooted in differences in the relative approach velocities, which will be studied next.
	
	\subsubsection{Relative velocities}
\begin{figure}
		\centerline{\includegraphics{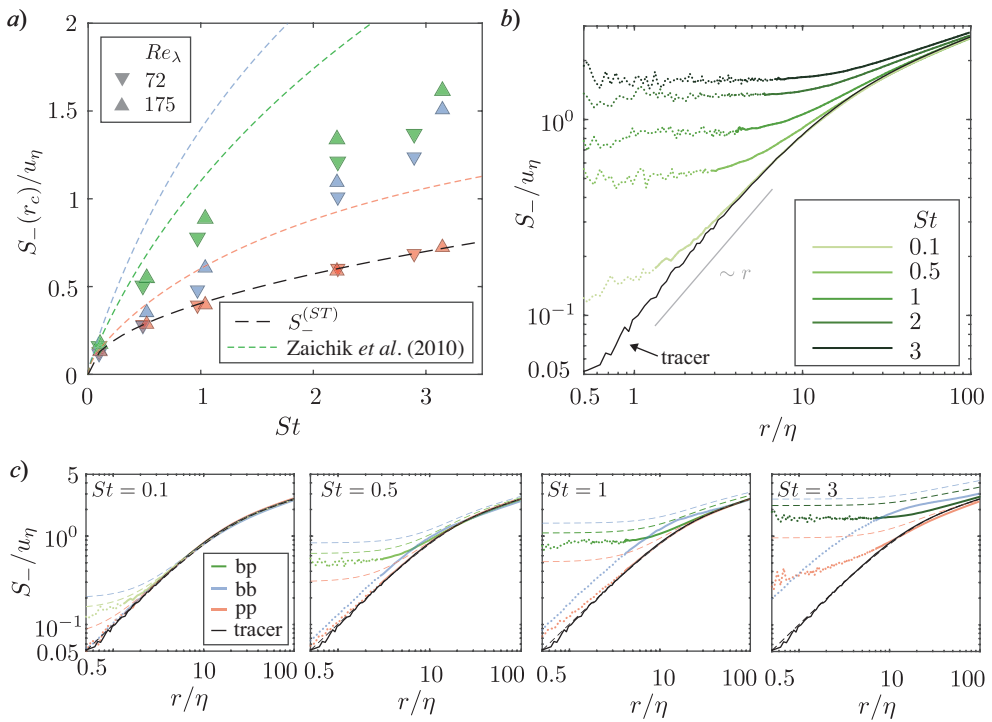}}
		\caption{(\textit{a}) The effective radial collision velocity. (\textit{b}) The effective bubble--particle radial approach velocity against pair separation $r$ \makeredTwo{at $\Rey_\lambda = 175$}. (\textit{c}) The effective radial approach velocity of each type of collisions across $St = 0.1$ to $St = 3$ \makeredTwo{at $\Rey_\lambda = 175$}. The dashed lines show the prediction by \citet{zaichik_turbulent_2010}.}
		\label{fig:velStat}
	\end{figure}

	We show results for the \makered{non-dimensional} effective radial approach velocity \makered{$S_-/u_\eta$} at collision distance in figure \ref{fig:velStat}(\textit{a}). The \makered{non-dimensional} approach velocity increases with $St$ and also slightly with $\Rey_\lambda$ for \makered{collisions involving bubbles, whereas it has only a very weak $\Rey_\lambda$ dependence for pp collisions, as also reported in \citet{ireland_effect_2016}}. Consistent with figure \ref{fig:collKShearOverCollK}, the approach velocities are highest for bp collisions and lowest in the pp case. 
	
	To understand these trends, it is instructive to consider $S_-$ as a function of $r$, which is presented in figure \ref{fig:velStat}(\textit{b}) for bp collisions for different $St$, and in the plots of figure \ref{fig:velStat}(\textit{c}) for different collision types at constant $St$.
	Also shown in these figures is the effective approach velocity for tracer particles, for which $S_-$ is proportional to $r$ in the dissipation range, as expected. In particular, for low $St$, where $S_-(r)$ remains close to the tracer curve, this in part explains the $St$ dependence of $S_-(r_c)$ as $r_c$ increases with increasing $St$. However, the curves also `peel off' the linear scaling at increasingly larger distances and to a larger extent as $St$ increases. For the pp case, this is a well-known manifestation of the `sling effect', where due to path history effects, particles arrive at the same location with different velocities such that their relative velocities exceed those of the fluid. Note that while the bb case has a larger $S_-$ than the tracer case, the linear scaling remains mostly unchanged there, such that the difference presumably is rather due to preferential concentration effects and the fact that the slip velocity at the same $St$ is higher for this case (see (\ref{eq:FouxonSlip})). The deviations from the tracer behaviour are strongest for the bp case throughout, and already for $St\geq 2$, $S_-$ are essentially independent of $r$ for $r\lessapprox 10\eta$. The decorrelation between bubble and particle velocities implicit in these observations is the reason why $S_-$ is largest for all $St$ in figure  \ref{fig:velStat}(\textit{a}).
	The model of \citet{zaichik_turbulent_2010} fails to reproduce this feature. Instead, $S_-$ from the model is highest for bb collisions, such that the trend falls in line with that of $v_i'^2$ (see figure \ref{fig:vrms}). Plotting the model prediction against the separation distance in figure \ref{fig:velStat}(\textit{c}) confirms that this trend persists over all $r$. Approach velocities are also generally overpredicted by the model, which is in part also related to the differences observed for $v_b'^2$ in figure \ref{fig:vrms} and to the significant peel off the linear scaling especially for the bb case. Consistent with the results in figure 7, $S_-^{pp}$ at both $\Rey_\lambda$ is matched very well by the Saffman--Turner model $ S_-^{(ST)} = 0.5r_c\sqrt{2\varepsilon/15\nu\pi}$, where the prefactor 0.5 accounts for the fact that only separating pairs are considered here.
	
\begin{figure}
		\centerline{\includegraphics{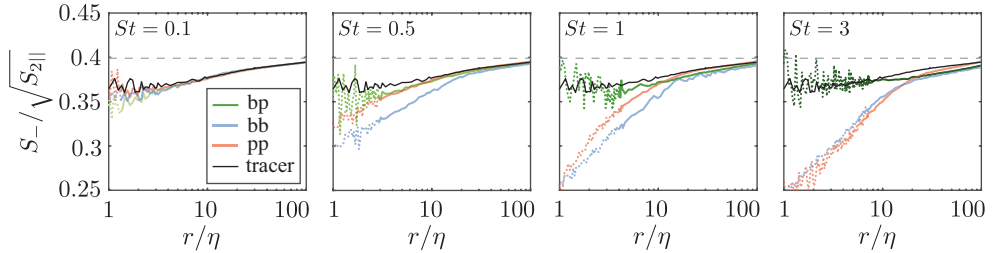}}
		\caption{The ratio of the effective radial approach velocity to the standard deviation of the radial component of the relative velocity $\Delta v_r$ \makeredTwo{at $\Rey_\lambda = 175$}.}
		\label{fig:Sratio}
	\end{figure}

	In modelling approaches \makered{\citep[e.g.][]{abrahamson_collision_1975,yuu_collision_1984,kruis_collision_1997,ngo-cong_isotropic_2018,zaichik_turbulent_2010}}, it is common to deduce $S_-$ based on a Gaussian distribution of the relative \makered{radial} velocity $\Delta v_r$. In this case, the ratio $S_-/\sqrt{S_{2\parallel}}$ can be determined to be $1/\sqrt{2\pi} \approx 0.4$, where the variance is defined as
	\begin{equation}    \label{eq:S2par}
	S_{2\parallel}(r) = \int_{-\infty}^{\infty} (\Delta v_r)^2\mathrm{p.d.f.}(\Delta v_r|r) \mathrm{d}(\Delta v_r).
	\end{equation}
	It was already reported in \citet{ireland_effect_2016} that $S_-/\sqrt{S_{2\parallel}}$ can drop significantly below the Gaussian value for pp collisions at $St \sim 1$ and for small $r_c$. This is confirmed by our results in figure \ref{fig:Sratio}, where also the bb case is seen to follow similar trends. For bp collisions, the ratio remains much closer to the Gaussian value, almost indifferent to $St$ and close to the tracer result. Relative velocities are therefore better approximated by the Gaussian assumption for the bp case, while the overprediction of $S_-$ resulting from doing so is larger for the monodisperse collisions. 
	
	\begin{figure}
		\centerline{\includegraphics{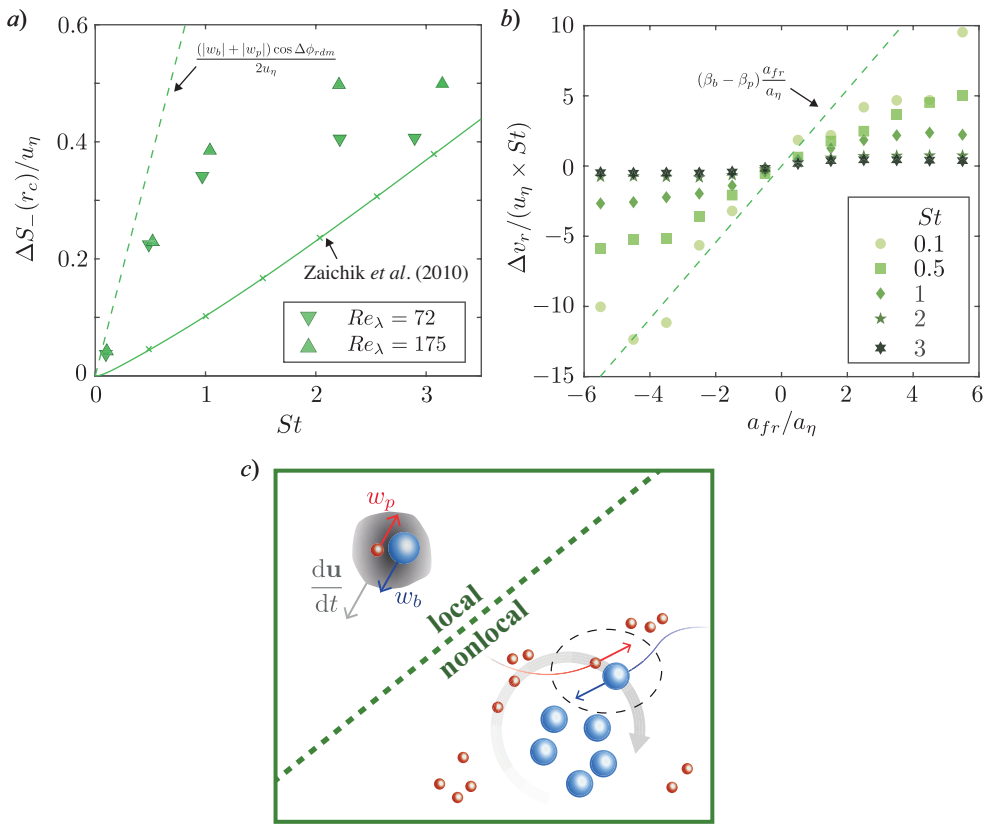}}
		\caption{(\textit{a}) The excess bubble--particle effective radial collision velocity $\Delta S_- = S_-^{bp} - (S_-^{bb}+S_-^{pp})/2$ from simulations over a range of $St$. The factor $1/2$ for the dashed line accounts for the fact that $S_-$ is obtained by averaging only over the negative portion of $\mathrm{p.d.f.}(\Delta v_r|r_c)$ as defined in (\ref{eq:Smin}). \makeredTwo{(\textit{b}) The radial component of the relative velocity conditioned on pairs with $r\in[r_c-\eta/2,r_c+\eta/2]$ binned by the local radial fluid acceleration at $\Rey_\lambda=175$. (\textit{c})} Schematic illustrations of the local and non-local turnstile mechanisms.}
		\label{fig:velCorrelation}
	\end{figure}
	
	For a more quantitative investigation of variation of $S_-$ across collision types, we plot the excess bubble--particle effective radial collision velocity $\Delta S_- = S_-^{bp} - (S_-^{bb}+S_-^{pp})/2$ in figure \ref{fig:velCorrelation}(\textit{a}). At small $St$, such an additional approach velocity arises for the bp case from the fact that, due to the change in sign of $\beta_i$ in (\ref{eq:FouxonSlip}), bubbles and particles react differently when subjected to the same fluid acceleration. \makeredTwo{Indeed, figure \ref{fig:velCorrelation}(\textit{b}) shows that $\Delta v_r^{bp}$ conditioned on pairs with $r\in[r_c-\eta/2,r_c+\eta/2]$ at small $St$ closely follows the resulting expression for the relative velocity $\Delta v_r/u_\eta = St(\beta_b - \beta_p)a_{fr}/a_\eta$, where $a_{fr}$ is the average radial fluid acceleration at bubble and particle positions.} We \makeredTwo{therefore }conceptualise \makeredTwo{the resulting excess bp collision velocity} as the local `turnstile mechanism', since bubbles and particles respectively move in opposite directions, as illustrated in \makeredTwo{figure \ref{fig:velCorrelation}(\textit{c})}. An estimate for this effect can be obtained using $a_f \sim a_\eta$ in (\ref{eq:FouxonSlip}) and assuming a random orientation between the particle separation vector and the fluid acceleration, resulting in an average angle $\phi_{rdm} = 1 \mathrm{rad} \approx 57^\circ$. This is shown by the dashed line in figure \ref{fig:velCorrelation}(\textit{a}). While not strictly a prediction for $\Delta S_-$, this estimate is largely consistent with the data for $St<1$, suggesting that the enhanced approach velocity in this range is indeed due to the local turnstile effect. \makeredTwo{This is further consistent with the fact that for $St\geq 1$, the correlation of $\Delta v_r^{bp}$ with the local fluid acceleration is seen to vanish almost completely in figure \ref{fig:velCorrelation}(\textit{b}).} A positive $\Delta S_-$ is also predicted by the model of \citet{zaichik_turbulent_2010}, but underestimates the value in the low-$St$ regime.
	Beyond $St = 1$, $\Delta S_-$ flattens off at high values. Along with the observations made in figure \ref{fig:velStat}(\textit{c}), this suggests that the `path history' effect active at these $St$ values is enhanced for the bp case. We can rationalise this in terms of a non-local `turnstile mechanism' (see \makeredTwo{figure \ref{fig:velCorrelation}\textit{c}}), where differences in the particle velocities arise as bubbles and particles interact differently with larger and more energetic eddies in the flow. As a consequence, their local velocities differ more than for monodisperse cases at equal $St$. Both the local and non-local inertial effects discussed here enhance the relative approach velocity and therefore explain why $\Gamma^{(STc)}_{bp} < \Gamma_{bp}$ as observed before.
	
	\subsection{Lagrangian statistics}\label{sec::LagStat}
	
	To better examine the collision process, we adopt the Lagrangian point of view, where colliding pairs are tracked individually for separations $r \leq 100\eta$ up to collision. The corresponding statistics averaging over all pairs with the same $r$ are denoted by the suffix $\mathcal{L}$. 
	
\begin{figure}
		\centerline{\includegraphics{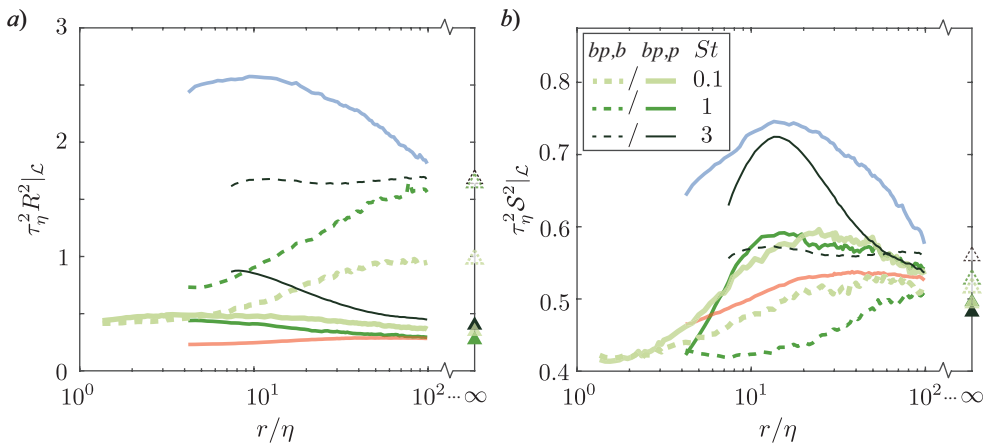}}
		\caption{Norm of (\textit{a}) the rotation rates and (\textit{b}) the strain rates of the flow at bubble and particle positions against pair separation conditioned on colliding pairs at $\Rey_\lambda = 175$ for $r \geq r_{c}$. Only the $St = 1$ case is shown for bubble--bubble and particle--particle pairs. Also plotted at $r\rightarrow+\infty$ is the unconditioned $R^2$ and $\mathcal{S}^2$ at the respective $St$.}
		\label{fig:LagStatRsqSsq}
	\end{figure}

	In \makered{figure \ref{fig:LagStatRsqSsq}}, we show how $R^2|_\mathcal{L}$ and $\mathcal{S}^2|_\mathcal{L}$ vary as functions of $r$ for the colliding pairs. Consistent with the results in figure \ref{fig:RsqSsq}, variations are significantly stronger for $R^2|_\mathcal{L}$ (figure \ref{fig:LagStatRsqSsq}\textit{a}) compared to $\mathcal{S}^2|_\mathcal{L}$ (figure \ref{fig:LagStatRsqSsq}\textit{b}). In contrast to the monodisperse case, bubbles and particles need to leave their preferred flow regions in order to collide. While at $r/\eta \sim 100$ the unconditioned preferential concentration (indicated by the symbols in figure \ref{fig:LagStatRsqSsq}) is mostly recovered, a pronounced drift generally sets in for $r/\eta \lessapprox 50$. This might suggest that interaction with eddies at this intermediate range plays a role in bringing bubbles and particles together from their segregated locations. At $St = 0.1$, bubbles and particles appear to occupy similar flow regions for $r/\eta \lessapprox 10$ before colliding, i.e. in particular, the $R^2|_\mathcal{L}$ curve flattens at these scales, whereas the same is not the case at $St = 1$ and $St = 3$, where the collision distance is larger. These observations are consistent with the considerations regarding the local and non-local turnstile mechanisms made above. It is noteworthy that $R^2|_\mathcal{L}$ at $St \leq 1$ varies mostly for bubbles as they approach particles, while at $St = 3$, $R^2|_\mathcal{L}$ is almost constant for the bubbles but varies more for the particles. The latter case also stands out in the $\mathcal{S}^2|_\mathcal{L}$ plot as the slight decrease in strain values towards collision is not observed there.
	
\begin{figure}
		\centerline{\includegraphics{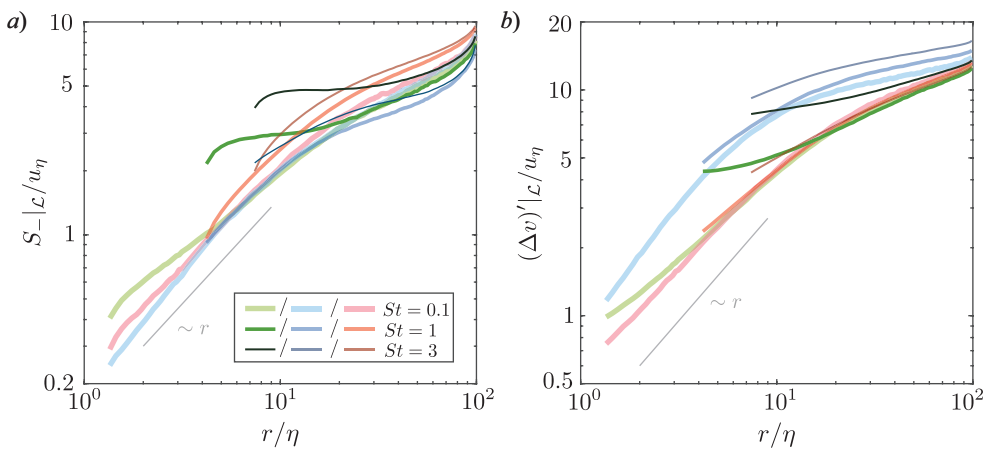}}
		\caption{(\textit{a}) The effective approach velocity, and (\textit{b}) the r.m.s. relative speed, conditioned on colliding pairs at $\Rey_\lambda = 175$.}
		\label{fig:LagStatVel}
	\end{figure}

	In figure \ref{fig:LagStatVel}, we present velocity statistics conditioned on the colliding pairs. In particular, we consider the approach velocity $S_-|_\mathcal{L}$ (figure \ref{fig:LagStatVel}\textit{a}) as well as the r.m.s. relative speed $(\Delta v)'|_\mathcal{L}$ (figure \ref{fig:LagStatVel}\textit{b}). The approach velocities differ slightly in magnitude from the unconditioned results in figure \ref{fig:velStat}, but the trends remain consistent. In particular, $S^{bp}_-|_\mathcal{L}$ is again largest for separations close to the respective collision distances for all $St$ shown. This is remarkable in view of the fact that $(\Delta v)'|_\mathcal{L}$ is actually largest for the bb case. This implies that $\boldsymbol{\Delta v}$ and $\boldsymbol{r}$ tend to be more anti-aligned for the bp case to yield the higher relative approach velocity.
	
	\begin{figure}
		\centerline{\includegraphics{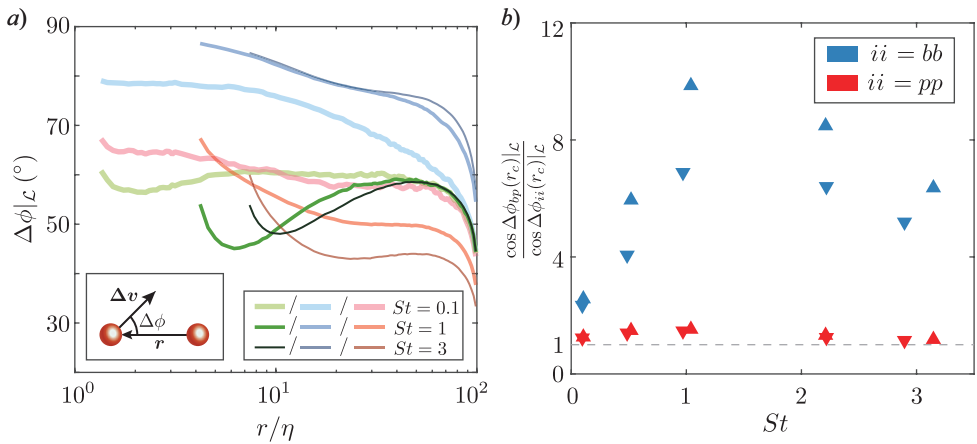}}
		\caption{(\textit{a}) The angle between the separation and relative velocity vectors $\Delta\phi$ for $r \geq r_{c}$ at $\Rey_\lambda = 175$. Bubble--particle/bubble--bubble/particle--particle pairs are represented by green/blue/red lines as in figure \ref{fig:LagStatVel}. \makered{The inset shows the definition of $\Delta\phi$ in the rest frame of the particle on the right.} (\textit{b}) The extra angular contribution to the radial component of the bubble--particle collision velocity relative to the bubble--bubble and particle--particle cases.}
		\label{fig:LagStatCollAngle}
	\end{figure}
	
	To measure this, the angle between the separation vector and the relative velocity vector $\Delta\phi$ \makered{(see the inset of figure \ref{fig:LagStatCollAngle}\textit{a}) is used. For a head-on approach, $\Delta\phi=0^\circ$, while $\Delta\phi = 90^\circ$ means that the other particle is circling around in the rest frame of the collision partner. Note that as colliding particles must be approaching each other, $0^\circ\leq \Delta\phi<90^\circ$.}
	As demonstrated in figure \ref{fig:LagStatCollAngle}(\textit{a}), the bubble--particle pairs are indeed the most anti-aligned, i.e. \makered{$\Delta\phi|_\mathcal{L}$} is lowest, relative to the other types of collisions near $r_{c}$.
	Figure \ref{fig:LagStatCollAngle}(\textit{b}) shows the ratio of the cosines of \makered{$\Delta\phi|_\mathcal{L}(r_c)$} as a measure for how much differences in alignment lead to a relative enhancement of $S_-(r_c)$. The ratio is up to 10 for bp relative to bb collisions, and up to $\sim1.5$ compared to particle--particle collisions, and values for both cases are slightly higher at the higher $\Rey_\lambda$. When interpreting the smaller difference between pp and bp collisions here, it should be kept in mind that based on the discussion around figure \ref{fig:collKShearOverCollK}, the collision mechanism differs between these two cases: whereas pp collisions appear predominantly shear-driven, where a good alignment may be expected, inertial effects, for which this is not necessarily the case, play a much more significant role for bp collisions. 
	
	\subsection{\makered{Effect of lift, finite particle density and nonlinear drag}} \label{sec::LiftDensityDrag}
\begin{figure}
		\centerline{\includegraphics[]{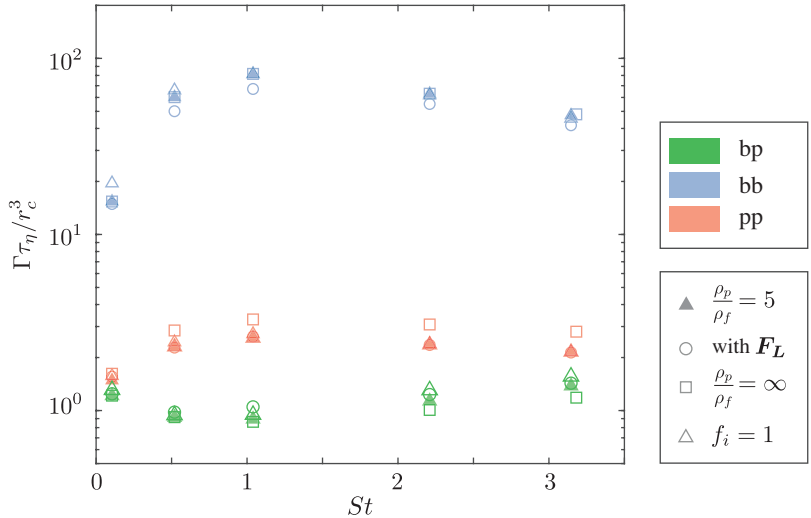}}
		\caption{The dimensionless collision kernel for different particle densities simulated, with the lift force included and for the case when the drag correction factor is $f_i = 1$.}
		\label{fig:rhopInfCollK}
	\end{figure}

\begin{figure}
		\centerline{\includegraphics{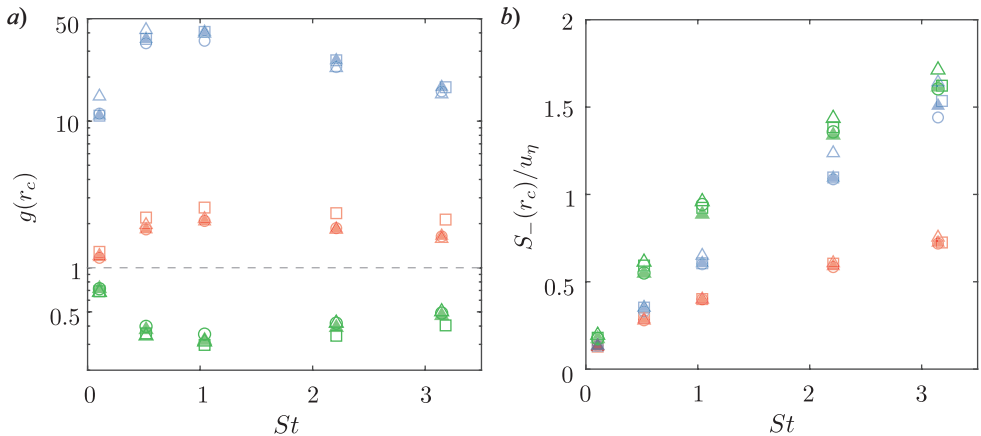}}
		\caption{(\textit{a}) The RDF at collision distance, and (\textit{b}) the effective radial collision velocity for different particle densities, with the lift force included and for the case when the drag correction factor is $f_i = 1$. The symbols follow figure \ref{fig:rhopInfCollK}.}
		\label{fig:rhopInfRDFVel}
	\end{figure}

	Our simulations are performed \makered{without the lift force using a finite particle density $\rho_p/\rho_f=5$} and a non-linear drag law \makered{with $f_i = 1+0.169Re_i^{2/3}$}. To test how sensitive our results are \makered{to these parameters, we ran addition simulations for $\Rey_\lambda=175$ changing one parameter at a time: adding the lift force $\boldsymbol{F_L} = -2/3\rho_f\pi r_i^3(\boldsymbol{v_i}-\boldsymbol{u})\times (\nabla\times\boldsymbol{u})$ to the right hand side of (\ref{eq:SimpleParticleEoM}), setting $\rho_p/\rho_f=\infty$, or using Stokes drag ($f_i = 1$) for both phases. For all of these simulations, $r_c$ is chosen to be identical to the case when $\rho_p/\rho_f=5$.} 
	We compare the results for $\Gamma$ in figure \ref{fig:rhopInfCollK}, and those for $g(r_{c})$ and $S_-(r_{c})$ in figure \ref{fig:rhopInfRDFVel}. \makered{The plots show that the collision statistics are not sensitive to the lift force. Also more generally, the effect of the lift force is limited, with the most noticeable change occurring for the preferential sampling where we observed a slight decrease in $\tau_\eta\langle R^2\rangle_r$ at bubble clusters, consistent with \citet{mazzitelli_effect_2003}. It is also} evident that the particle density affects these quantities only marginally. Similarly, \makered{the results remain largely unchanged for the $f_i = 1$ case}, with the exception of a slight increase in $\Gamma_{bp}$ due to a higher $S_-$ at the larger $St$. \makered{We furthermore note that other expressions for $f_i$ have been proposed in the literature \citep{schiller_uber_1933,wan_study_2020,clift_bubbles_2005}. These are compared in \makered{Appendix \ref{sec::DragCorrection}}, which shows that they do not differ significantly when $Re_i \lesssim 100$. Hence we conclude that our results would not be sensitive to whichever one of these parametrisations is chosen.} Based on the findings in this section, it therefore appears appropriate to \makered{neglect lift, and to} use the simplifications of linear drag and \makered{$\rho_p/\rho_f \to \infty$ for simulations and modelling approaches} in the present parameter range.
	
	\section{Discussion and conclusion}  \label{sec::discussion&conclusion}
\begin{figure}
		\centerline{\includegraphics{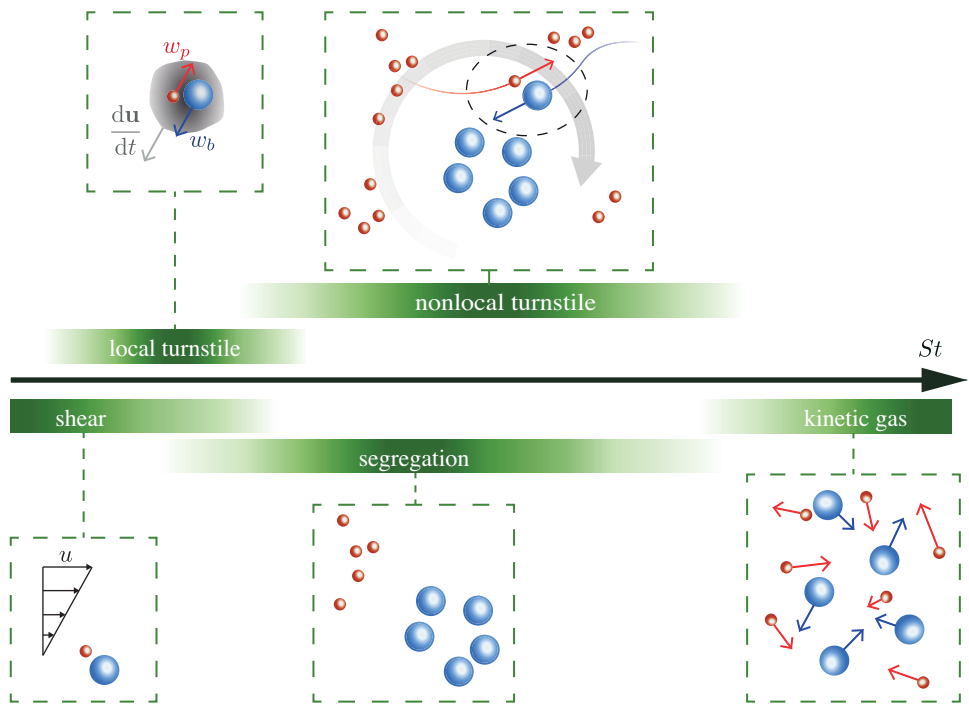}}
		\caption{A sketch of the bubble--particle collision mechanisms with their approximate $St$ dependence.}
		\label{fig:bpoverview}
	\end{figure}

	We studied bubble--particle collisions in turbulence using a point-particle approach. \makered{Besides a critical appraisal of current models for the problem, our results highlight that the difference in the density ratios between bubbles and particles critically affects the collision statistics}. 
	An overview of the physical picture that emerges is presented in figure \ref{fig:bpoverview}. For $St \to 0$, the shear mechanism is applicable, \makered{meaning that the collision velocity is well-described by the local fluid velocity gradient}. Once $St$ reaches finite values, bubbles and particles concentrate preferentially in different regions of the flow, which leads to their segregation. This effect reduces the \makered{bubble--particle collision kernel $\Gamma_{bp}$ and} is maximal for $St \sim 1$. The correction remains $\textit{O}(1)$, which is comparable to the enhancement of the collision kernel due to clustering for \makered{particle--particle} cases, but much less than that for the \makered{bubble--bubble} cases, respectively.  For \makered{particle--particle} collisions (evaluated at $r_c = r_b + r_p$), the relative approach velocities are found to be entirely consistent with the shear mechanism for the full range of $St$ explored here, such that the collision kernel in these cases is very well approximated by \makered{extending the shear-induced collision kernel to account for non-uniform particle concentration} $\Gamma^{(STc)}$. The same does not hold for \makered{bubble--particle} and \makered{bubble--bubble} collisions, where inertial effects additionally play a role and enhance the effective relative approach velocities beyond the shear scaling. This enhancement is strongest for the \makered{bubble--particle} case, and we conceptualise this effect in terms of a `turnstile' mechanism, which is related to the fact that fluid accelerations lead to opposing drift velocities relative to the fluid for bubbles and particles. This happens locally, i.e. on the same fluid element, if $St$ is sufficiently small for bubbles/particles to follow pathlines ($St\leq 0.5$ based on the results in figure \ref{fig:velCorrelation}), but also appears to play a role at higher $St$, where the path history becomes more relevant. \makered{For $St \gtrapprox 2$, the ratio $\Gamma^{(STc)}/\Gamma$ and the values of the effective radial approach velocity at contact $S_-(r_c)$} are comparable for \makered{bubble--particle} and \makered{bubble--bubble} collisions, which indicates that the `non-local' turnstile is less prevalent there, in line with the fact that local particle/bubble velocities become increasingly random at larger $St$.
	
	Our method does not allow us to assess the high $St$ (kinetic-gas-like) regime since bubble sizes violate the point-particle assumption. Applying the \citet{abrahamson_collision_1975} model pertinent to this regime to the present data vastly overpredicts the collision kernel, and the apparent improvement by adaptations of this framework is merely an artefact of using inconsistent velocity expressions. Our data for bubble--particle collisions are better approximated by the models of \citet{saffman_collision_1956} and \citet{zaichik_turbulent_2010} when taking an additional correction for the inhomogeneous distribution into account. The predictions based on the theory of \citet{zaichik_turbulent_2010} generally overestimate $S_-$ and fail to reproduce the finding that $S_-^{bp}$ is largest in the present parameter range when comparing across species. These observations are not affected when using a linear drag relation, i.e. $f_i = 1$ in (\ref{eq:SimpleParticleEoM}), or in the limit $\rho_p/\rho_f \to \infty$ as shown in \makered{\S \ref{sec::LiftDensityDrag}}. We therefore could not reproduce the good agreement with the \citet{zaichik_turbulent_2010} theory for $S_-^{bp}$ reported in \citet{fayed_direct_2013}. Potentially, this is due to different choices for $St_b$, which we set to equal to $St_p$, whereas this parameter is kept constant in \citet{fayed_direct_2013}.
	
	In our simulations, we employ the non-dimensional control parameters $St$ and $\Rey_\lambda$. In practice, the fluid and particle properties are controlled separately so using $r_b$, $r_p$ and $\varepsilon$ is more common. However, by definition, changing $\varepsilon$ affects both $St$ and $\Rey_\lambda$. To assess the overall effect, consider the region between $St = 0.1$ and $St = 0.5$, where \makered{the non-dimensional} collision kernel $\Gamma_{bp}\tau_\eta/r_c^3$ decreases most sharply and drops by a factor of about 1.3.
	As the particle parameters are constant, $\tau_i = \textrm{const.}$, this is associated with a 5-fold decrease in $\tau_\eta$. According to the shear scaling, which appears to be applicable to our data, this increase in turbulence intensity implies also a 5-fold increase in $\Gamma_{bp}$. Hence there remains a net benefit from increasing the turbulence intensity even in this range, although the effect is reduced somewhat due to segregation.
	
	Our results show that considering finite \makered{particle density, nonlinear drag and the lift force} have limited effect on the bubble--particle collision rate. However, other relevant factors are yet to be explored. Most importantly, this concerns the \makered{buoyancy force and finite size} effects in particular for the bubble motion and non-identical $St$ for bubbles and particles, as well as modelling the collision behaviour of bubbles and particles.
	\makered{More complex simulations as well as experiments are certainly needed to make precise predictions of the bubble-particle collision rate in realistic settings. However, as far as modelling approaches and a physical understanding are concerned, we believe that the most promising approach to disentangle the effect of additional factors (such as buoyancy) is to treat them as modifications to a simpler base case like the one studied here. }
	
	\section*{Acknowledgements}
	We thank Rodolfo Ostilla-M\'{o}nico \makered{and Linfeng Jiang} for useful discussions as well as Rui Yang, Sreevanshu Yerragolam and Chris Howland for help with visualisation and code compilation.
	
	\section*{Funding}
	This project has received funding from the European Research Council (ERC) under the European Union's Horizon 2020 research and innovation programme (grant agreement No. 950111, BU-PACT). The simulations are conducted with the Dutch National Supercomputers Cartesius and Snellius, and MareNostrum4 of the Barcelona Supercomputing Center.
	
	\section*{Declaration of interests}
	The authors report no conflict of interest.
	
	\section*{Data availability statement}
	All data supporting this study are openly available from the 4TU.ResearchData repository at \href{https://doi.org/10.4121/22147319}{https://doi.org/10.4121/22147319}.
	
	\appendix
	\section{Extension of the \citet{kruis_collision_1997} model to particles with different densities}    \label{sec::extKruisKusters}
	
	\makered{The original model by \citet{kruis_collision_1997} is applicable to collisions of particles with arbitrary but equal density only. Here, we extend their framework to collisions of particles with different densities.
		In their model, \citet{kruis_collision_1997} considered two limits: small and large $St$. We follow their approach to derive the corresponding expressions for particles with different densities. For convenience, we first rewrite (\ref{eq:SimpleParticleEoM}) as
		\begin{equation}    \label{eq:SimpleParticleEoM_gamma}
		\frac{d\boldsymbol{v_i}}{dt} = \gamma_i\frac{\mathrm{D}\boldsymbol{u}}{\mathrm{D}t} + \frac{\boldsymbol{u}-\boldsymbol{v_i}}{\tau_i},
		\end{equation}
		where $\gamma_i = \frac{3\rho_f}{2\rho_i + \rho_f}$ and $i = 1,2$ as in \S\ref{sec::intro}.}
	
	\subsection{Small $St$ limit}   \label{sec::extKruisKusters_smallSt}
	\makered{For small $St$, \citet{kruis_collision_1997} based their model on \citet{yuu_collision_1984}, which considers a `local' inertial effect due to the different response to fluid fluctuations of particles having different $St$ (originally termed the `accelerative mechanism' in \citet{kruis_collision_1997}) and the shear mechanism. The main contribution by \citet{kruis_collision_1997} is employing a more accurate expression of the fluid Eulerian energy spectrum that also captures the dissipation range behaviour
		\begin{equation}    \label{eq:fullEspec}
		E_f(\omega) = \frac{4u'^2\xi}{2\pi(\xi-1)}\bigg(\frac{T_{fL}}{1+T_{fL}^2\omega^2} - \frac{T_{fL}}{\xi^2 + T_{fL}^2\omega^2} \bigg),
		\end{equation}
		where $\omega$ is the angular frequency,
		\begin{equation}
		\xi = 2\bigg(\frac{T_{fL}}{\tau_L}\bigg)^2,
		\end{equation}
		$T_{fL}=0.4L_f/u'$ is the fluid Lagrangian integral time scale, $L_f$ is the Eulerian longitudinal integral length scale, and $\tau_L^2=2u'^2/\langle(\mathrm{D}u/\mathrm{D}t)^2\rangle = 2u'^2/(1.16\varepsilon^{3/2}\nu^{-1/2})$ is the square of the Lagrangian time scale.}
	
	\makered{For the local inertial effect, the ensemble average of the square of the collision velocity is
		\begin{equation}
		\langle[(\Delta v)^{(loc)}_{12}]^2\rangle = v_1'^2 + v_2'^2 - 2\langle v_1 v_2\rangle
		\end{equation}
		where $\langle\cdot\rangle$ denotes ensemble averaging, and 
		\begin{equation}
		v_i^2 = u'^2\frac{\xi}{\xi-1}\bigg[\frac{1 + \gamma_i^2 \tau_i/T_{fL}}{1 + \tau_i/T_{fL}} - \frac{1 + \gamma_i^2\xi\tau_i/T_{fL}}{\xi(1+\xi\tau_i/T_{fL})}\bigg]
		\end{equation}
		as derived in \citet{kruis_collision_1997}. The integral giving the particle velocity correlation term follows \citet{yuu_collision_1984}, whose model has been extended to particles with unequal densities in \citet{ngo-cong_isotropic_2018}. The integral in \citet{ngo-cong_isotropic_2018} reads
		\begin{equation}    \label{eq:vCorrLI_integral}
		\langle v_1 v_2\rangle = \int_{0}^{+\infty} \frac{(\tau_1^{-2}+\gamma_1 \omega ^2) (\tau_2^{-2}+\gamma_2 \omega
			^2)+\tau_1^{-1} \tau_2^{-1} (\gamma_1-1) (\gamma_2-1) \omega^2}{(\tau_1^{-2}+\omega ^2) (\tau_2^{-2}+\omega^2)} E_f(\omega) d\omega
		\end{equation}
		Note that although \citet{kruis_collision_1997} found a misprint in the integral given by \citet{yuu_collision_1984}, this error has not propagated to \citet{ngo-cong_isotropic_2018}. Performing the integration yields
		\begin{eqnarray}    \label{eq:vCorrLI}
		\frac{\langle v_1 v_2 \rangle}{u'^2} & = & 1 + \frac{\xi}{(\xi-1)(\tau_1^{-1} + \tau_2^{-1})}\bigg[\frac{(\gamma_1-1)(\tau_2^{-1}+\tau_1^{-1}\gamma_2)}{1+\tau_1^{-1}T_{fL}} + \frac{(\tau_1^{-1} + \tau_2^{-1}\gamma_1)(\gamma_2-1)}{1+\tau_2^{-1}T_{fL}} \nonumber\\ && - \frac{(\gamma_1-1)(\tau_2^{-1} + \tau_1^{-1}\gamma_2)}{\tau_1^{-1}T_{fL} + \xi} - \frac{(\tau_1^{-1} + \tau_2^{-1}\gamma_1)(\gamma_2 - 1)}{\tau_2^{-1}T_{fL} + \xi} \bigg],
		\end{eqnarray}
		which is different from the corresponding expression in \citet{ngo-cong_isotropic_2018} solely because of the usage of another form of $E_f(\omega)$ as given by (\ref{eq:fullEspec}) in (\ref{eq:vCorrLI_integral}).
	}
	
	\makered{For the collision velocity due to the shear mechanism, \citet{kruis_collision_1997} used the result in \citet{yuu_collision_1984}:
		\begin{equation}
		\langle[(\Delta v)^{(shr)}_{12}]^2\rangle = \frac{\varepsilon}{5\nu}\bigg(\frac{v_1'^2}{u'^2}r_1^2 + \frac{v_2'^2}{u'^2}r_2^2 + 2\frac{\langle v_1 v_2 \rangle}{u'^2}r_1r_2\bigg).
		\end{equation}
		Using (\ref{eq:StokesNumber}) and allowing $\gamma_1 \neq \gamma_2$ gives
		\begin{equation}
		\langle[(\Delta v)^{(shr)}_{12}]^2\rangle = \frac{3\varepsilon}{5}\bigg(\frac{v_1'^2}{u'^2}\frac{\tau_1}{\gamma_1} + \frac{v_2'^2}{u'^2}\frac{\tau_2}{\gamma_2} + 2\frac{\langle v_1 v_2 \rangle}{u'^2}\sqrt{\frac{\tau_1\tau_2}{\gamma_1\gamma_2}}\bigg).
		\end{equation}
	}
	
	\makered{The overall collision kernel is then given by
		\begin{equation}    \label{eq:extendKruisKustersGammaSmallSt}
		\Gamma_{12} = \sqrt{8\pi}r_c^2\sqrt{\langle[(\Delta v)^{(loc)}_{12}]^2\rangle + \langle[(\Delta v)^{(shr)}_{12}]^2\rangle}.
		\end{equation}
		Note that in contrast to the expression in \citet{kruis_collision_1997}, (\ref{eq:extendKruisKustersGammaSmallSt}) does not carry a factor $1/\sqrt{3}$ as $u'$ is defined as the single-component r.m.s. fluid velocity.}
	
	\subsection{Large $St$ limit}
	\makered{The \citet{kruis_collision_1997} expression of $\Gamma$ for large $St$ is essentially based on \citet{williams_particle_1983} except that the added mass term is retained. As in \S\ref{sec::extKruisKusters_smallSt}, the densities of the different species $\gamma_i$ are assumed to be distinct. In this limit, the shear mechanism is not considered, so
		\begin{equation}
		\langle(\Delta v)^2_{12}\rangle = v_1'^2 + v_2'^2 - 2\langle v_1 v_2\rangle
		\end{equation}
		and
		\begin{equation}
		\Gamma_{12} = \sqrt{8\pi}r_c^2\sqrt{\langle(\Delta v)^2_{12}\rangle}.
		\end{equation}
		Furthermore, a simpler version of $E_f(\omega)$, namely
		\begin{equation}
		E_f(\omega) = \frac{4u'^2}{2\pi}\bigg(\frac{T_{fL}}{1+T_{fL}^2\omega^2}\bigg)
		\end{equation}
		is used, which corresponds to an exponentially decaying fluid velocity autocorrelation function ``moving with the mean flow'' $R_E^*(t) = \exp({-|t|/T_{fL}})$ that does not account for the dissipation range. \citet{kruis_collision_1997} showed that
		\begin{equation}
		v_i'^2 = u'^2\frac{1 + \gamma_i^2\tau_i/T_{fL}}{1 + \tau_i/T_{fL}},
		\end{equation}
		which leaves only the particle velocity correlation to be determined.}
	
	\makered{To determine $\langle v_1v_2\rangle$, we integrate (\ref{eq:SimpleParticleEoM_gamma}) for $t\in(-\infty,0]$ and exercise the freedom that $t = 0$ can be chosen at any instant, to obtain
		\begin{equation}
		v_{ij}(\boldsymbol{x_i},t) = \gamma_i u_{ij}(\boldsymbol{x_i},t) + \frac{1-\gamma_i}{\tau_i}\int_{0}^{+\infty}u_{ij}(\boldsymbol{x_i},t-\phi)\exp\bigg(-\frac{\phi}{\tau_i}\bigg)\mathrm{d}\phi
		\end{equation}
		where $j = x,y,z$ are the individual components of the corresponding vector, and $\boldsymbol{x_i}$ is the position vector. Then using the approximation \citep{williams_particle_1983}
		\begin{equation}
		\langle u(\boldsymbol{x_1},t')u(\boldsymbol{x_2},t'') \rangle \approx R_E^*(t'-t'')\int_{0}^{+\infty}E_f(\omega)\cos{\bigg[\omega \widehat{\Delta v}\bigg(t - \frac{t'+t''}{2}\bigg)\bigg]}d\omega
		\end{equation}
		where $\widehat{\Delta v}$ is a non-dimensional particle relative velocity, we have
		\begin{eqnarray}
		\langle v_1 v_2 \rangle & = & \gamma_1\gamma_2u'^2 + \frac{(1-\gamma_1)(1-\gamma_2)}{\tau_1\tau_2} \int_{0}^{\infty}\int_{0}^{\infty}\int_{0}^{\infty}\exp\bigg(-\frac{|\psi-\phi|}{T_{fL}}\bigg)\frac{2}{\pi}u'^2\frac{T_{fL}}{1+\omega^2T_{fL}^2}  \nonumber\\ && 
		\times\cos{\bigg[\omega \widehat{\Delta v}\bigg(\frac{\psi+\phi}{2}\bigg) \bigg]} \exp\bigg(-\frac{\psi}{\tau_1} - \frac{\phi}{\tau_2}\bigg)\mathrm{d}\omega \mathrm{d}\psi \mathrm{d}\phi + \frac{(1-\gamma_1)\gamma_2}{\tau_1} \nonumber \\ &&
		\times \int_{0}^{\infty}\int_{0}^{\infty}\exp\bigg(-\frac{|\psi|}{T_{fL}}\bigg)\frac{2}{\pi}u'^2\frac{T_{fL}}{1+\omega^2T_{fL}^2}\cos\bigg({\omega}\widehat{\Delta v}\frac{\psi}{2}\bigg)\exp\bigg(-\frac{\psi}{\tau_1}\bigg) \mathrm{d}\omega \mathrm{d}\psi \nonumber\\ &&
		+ \frac{\gamma_1(1-\gamma_2)}{\tau_2}\int_{0}^{\infty}\int_{0}^{\infty}\exp\bigg(-\frac{|\phi|}{T_{fL}}\bigg)\frac{2}{\pi}u'^2\frac{T_{fL}}{1+\omega^2T_{fL}^2}\cos\bigg({\omega}\widehat{\Delta v}\frac{\phi}{2}\bigg) \nonumber \\ &&
		\times \exp\bigg(-\frac{\phi}{\tau_2}\bigg) \mathrm{d}\omega \mathrm{d}\phi.
		\end{eqnarray}
		As $\tau_1,\tau_2 \gg 1$, $\exp(-\psi/\tau_1 - \phi/\tau_2) \approx 0$ except when $\psi = \phi$. Hence we retain this case only in the arguments of cosine and $\exp(-\psi/\tau_1 - \phi/\tau_2)$. Furthermore, we replace $\widehat{\Delta v}$ by $\widehat{\Delta v_X} = \sqrt{(v_1'^2 + v_2'^2)/u'^2}$ as the particle velocity correlation should be weak at large $St$. Thus
		\begin{eqnarray}
		\frac{\langle v_1 v_2 \rangle}{u'^2} & = & \gamma_1\gamma_2 + \frac{(1-\gamma_1)(1-\gamma_2)}{\tau_1\tau_2} \int_{0}^{\infty}\int_{0}^{\infty}\int_{0}^{\infty}\exp\bigg(-\frac{|\psi-\phi|}{T_{fL}}\bigg) \mathrm{d}\psi \frac{2T_{fL}}{\pi(1+\omega^2T_{fL}^2)}  \nonumber\\ && 
		\times\cos{(\omega \widehat{\Delta v_X}\phi)} \exp\bigg[-\phi\bigg(\frac{1}{\tau_1} + \frac{1}{\tau_2}\bigg)\bigg] \mathrm{d}\omega \mathrm{d}\phi + \frac{(1-\gamma_1)\gamma_2}{\tau_1} \nonumber\\ &&
		\times \int_{0}^{\infty}\int_{0}^{\infty}\exp\bigg(-\frac{|\psi|}{T_{fL}}\bigg)\frac{2T_{fL}}{\pi(1+\omega^2T_{fL}^2)}\cos\bigg(\omega\widehat{\Delta v_X}\frac{\psi}{2}\bigg)\exp\bigg(-\frac{\psi}{\tau_1}\bigg) d\omega d\psi \nonumber\\ &&
		+ \frac{\gamma_1(1-\gamma_2)}{\tau_2}\int_{0}^{\infty}\int_{0}^{\infty}\exp\bigg(-\frac{|\phi|}{T_{fL}}\bigg)\frac{2T_{fL}}{\pi(1+\omega^2T_{fL}^2)}\cos\bigg(\omega \widehat{\Delta v_X}\frac{\phi}{2}\bigg) \nonumber\\ &&
		\times \exp\bigg(-\frac{\phi}{\tau_2}\bigg) \mathrm{d}\omega \mathrm{d}\phi.
		\end{eqnarray}
		Integrating first over $\omega$ and then the remaining variables finally results in
		\begin{eqnarray}
		\frac{\langle v_1 v_2 \rangle}{u'^2} & = & T_{fL}\frac{(1-\gamma_1)(1-\gamma_2)}{\tau_1\tau_2} \bigg[\frac{2}{\tau_1^{-1} + \tau_2^{-1} + \widehat{\Delta v_X}T_{fL}^{-1}} - \bigg(\frac{1}{\tau_1} + \frac{1}{\tau_2} + \frac{\widehat{\Delta v_X} + 1}{T_{fL}} \bigg)^{-1} \bigg]  \nonumber\\ && 
		+ \frac{(1-\gamma_1)\gamma_2}{\tau_1}\bigg(\frac{1}{T_{fL}} + \frac{1}{\tau_1} + \frac{\widehat{\Delta v_X}}{2T_{fL}} \bigg)^{-1} + \frac{\gamma_1(1-\gamma_2)}{\tau_2}\bigg(\frac{1}{T_{fL}} + \frac{1}{\tau_2} + \frac{\widehat{\Delta v_X}}{2T_{fL}} \bigg)^{-1}  \nonumber\\ && 
		+ \gamma_1\gamma_2.
		\end{eqnarray}}
	
	\begin{figure}
		\centering
		\includegraphics{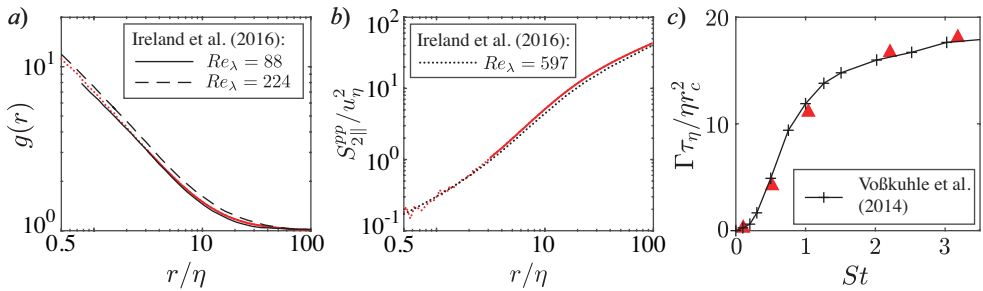}
		\caption{\makered{Particle--particle collision statistics of infinitely heavy particles: (a) the RDF $g_{pp}(r)$ at $St = 1$; (b) the variance of the radial component of the relative velocity $S_{2\parallel}^{pp}(r)$ at $St = 1$; and (c) the collision kernel $\Gamma_{pp}$ when taking $\rho_p/\rho_f=250$ and $r_c = 2r_p$.}}
		\label{fig:vosskuhleIrelandCompare}
	\end{figure}
	
	\section{Verification of point-particle code}  \label{sec::AppCodeVerification}
	\makered{To verify our code for the suspended phase, we compared the results of the infinitely heavy particle cases to the literature. As demonstrated by figure \ref{fig:vosskuhleIrelandCompare}, the RDF $g(r)$, the variance of the radial component of the relative velocity $S_{2\parallel}^{pp}(r)$ as defined by (\ref{eq:S2par}), and the collision kernel $\Gamma$ agree very well with data from \citet{ireland_effect_2016} and \citet{voskuhle_prevalence_2014}, especially considering the difference in $Re_\lambda$ compared to the literature data in figures \ref{fig:vosskuhleIrelandCompare}(\textit{a},\textit{b}).}
	
	\section{Comparison of different drag parametrisations} \label{sec::DragCorrection}
	
\begin{figure}
		\centerline{\includegraphics[]{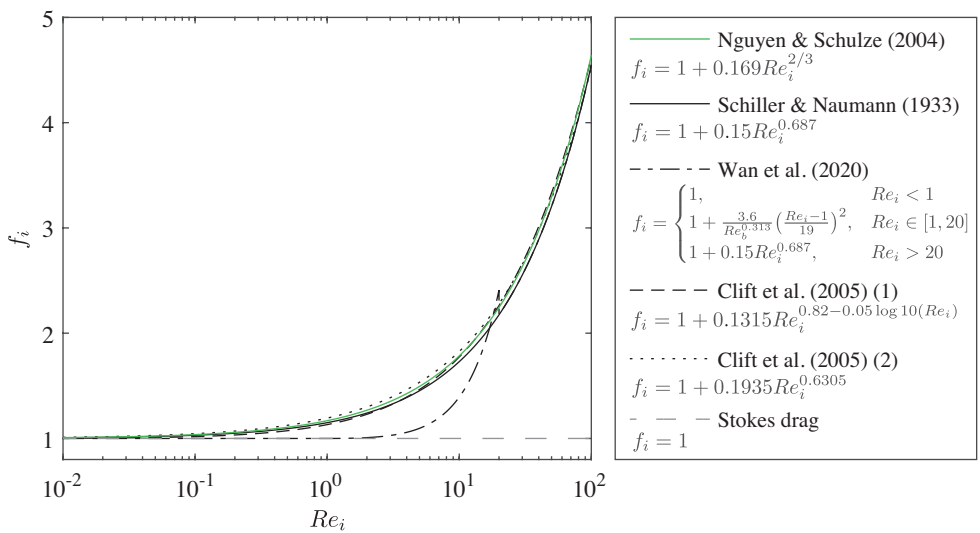}}
		\caption{\makered{Various drag correction factors $f_i$ from the literature compared to the one by \citet{nguyen_colloidal_2004} used in this study (solid green line).}}
		\label{fig:dragCorrvsRe}
	\end{figure}

	\makered{Various expressions of the drag correction factor $f_i$ have been proposed in the literature. Figure \ref{fig:dragCorrvsRe} shows that they are mostly similar for $Re_i \lessapprox 100$.}
	
	\pagebreak
	\bibliographystyle{jfm}
	\bibliography{bib_BuPaCT_PP_noGravity}

\begin{thebibliography}{73}
\expandafter\ifx\csname natexlab\endcsname\relax\def\natexlab#1{#1}\fi
\def\au#1{#1} \def\ed#1{#1} \def\yr#1{#1}\def\at#1{#1}\def\jt#1{\textit{#1}}
  \def\bt#1{#1}\def\bvol#1{\textbf{#1}} \def\vol#1{#1} \def\pg#1{#1}
  \def\publ#1{#1}\def\arxiv#1{#1}\def\org#1{#1}\def\st#1{\textit{#1}}

\bibitem[Abrahamson(1975)]{abrahamson_collision_1975}
{\sc \au{Abrahamson, J.}} \yr{1975}  \at{Collision rates of small particles in
  a vigorously turbulent fluid}.  \jt{Chem. Eng. Sci.}  \bvol{30}~(11),
  \pg{1371--1379}.

\bibitem[Aliseda {\em et~al.\/}(2002)Aliseda, Cartellier, Hainaux \&
  Lasheras]{aliseda_effect_2002}
{\sc \au{Aliseda, A.}, \au{Cartellier, A.}, \au{Hainaux, F.} \& \au{Lasheras,
  J.~C.}} \yr{2002}  \at{Effect of preferential concentration on the settling
  velocity of heavy particles in homogeneous isotropic turbulence}.  \jt{J.
  Fluid Mech.}  \bvol{468},  \pg{77--105}.

\bibitem[Bec {\em et~al.\/}(2007)Bec, Biferale, Cencini, Lanotte, Musacchio \&
  Toschi]{bec_heavy_2007}
{\sc \au{Bec, J.}, \au{Biferale, L.}, \au{Cencini, M.}, \au{Lanotte, A.},
  \au{Musacchio, S.} \& \au{Toschi, F.}} \yr{2007}  \at{Heavy particle
  concentration in turbulence at dissipative and inertial scales}.  \jt{Phys.
  Rev. Lett.}  \bvol{98}~(8),  \pg{084502}.

\bibitem[Bec {\em et~al.\/}(2005)Bec, Celani, Cencini \&
  Musacchio]{bec_clustering_2005}
{\sc \au{Bec, J.}, \au{Celani, A.}, \au{Cencini, M.} \& \au{Musacchio, S.}}
  \yr{2005}  \at{Clustering and collisions of heavy particles in random smooth
  flows}.  \jt{Phys. Fluids}  \bvol{17}~(7),  \pg{073301}.

\bibitem[Bewley {\em et~al.\/}(2013)Bewley, Saw \&
  Bodenschatz]{bewley_observation_2013}
{\sc \au{Bewley, G.~P.}, \au{Saw, E.-W.} \& \au{Bodenschatz, E.}} \yr{2013}
  \at{Observation of the sling effect}.  \jt{New J. Phys.}  \bvol{15},
  \pg{083051}.

\bibitem[Bloom \& Heindel(2002)]{bloom_structure_2002}
{\sc \au{Bloom, F.} \& \au{Heindel, T.~J.}} \yr{2002}  \at{On the structure of
  collision and detachment frequencies in flotation models}.  \jt{Chem. Eng.
  Sci.}  \bvol{57}~(13),  \pg{2467--2473}.

\bibitem[Brandt \& Coletti(2022)]{brandt_particle-laden_2022}
{\sc \au{Brandt, L.} \& \au{Coletti, F.}} \yr{2022}  \at{Particle-laden
  turbulence: progress and perspectives}.  \jt{Annu. Rev. Fluid Mech.}
  \bvol{54}~(1),  \pg{159--189}.

\bibitem[Calzavarini {\em et~al.\/}(2008{\natexlab{{\em a\/}}})Calzavarini,
  Cencini, Lohse \& Toschi]{calzavarini_quantifying_2008-1}
{\sc \au{Calzavarini, E.}, \au{Cencini, M.}, \au{Lohse, D.} \& \au{Toschi, F.}}
  \yr{2008{\natexlab{{\em a\/}}}}  \at{Quantifying turbulence-induced
  segregation of inertial particles}.  \jt{Phys. Rev. Lett.}  \bvol{101}~(8),
  \pg{084504}.

\bibitem[Calzavarini {\em et~al.\/}(2008{\natexlab{{\em b\/}}})Calzavarini,
  Kerscher, Lohse \& Toschi]{calzavarini_dimensionality_2008}
{\sc \au{Calzavarini, E.}, \au{Kerscher, M.}, \au{Lohse, D.} \& \au{Toschi,
  F.}} \yr{2008{\natexlab{{\em b\/}}}}  \at{Dimensionality and morphology of
  particle and bubble clusters in turbulent flow}.  \jt{J. Fluid Mech.}
  \bvol{607},  \pg{13--24}.

\bibitem[Chen {\em et~al.\/}(2006)Chen, Goto \&
  Vassilicos]{chen_turbulent_2006}
{\sc \au{Chen, L.}, \au{Goto, S.} \& \au{Vassilicos, J.~C.}} \yr{2006}
  \at{Turbulent clustering of stagnation points and inertial particles}.
  \jt{J. Fluid Mech.}  \bvol{553},  \pg{143}.

\bibitem[Chouippe \& Uhlmann(2015)]{chouippe_forcing_2015}
{\sc \au{Chouippe, A.} \& \au{Uhlmann, M.}} \yr{2015}  \at{Forcing homogeneous
  turbulence in direct numerical simulation of particulate flow with interface
  resolution and gravity}.  \jt{Phys. Fluids}  \bvol{27}~(12),  \pg{123301}.

\bibitem[Clift {\em et~al.\/}(2005)Clift, Grace \& Weber]{clift_bubbles_2005}
{\sc \au{Clift, R.}, \au{Grace, J.} \& \au{Weber, M.~E.}} \yr{2005} {\em
  Bubbles, drops, and particles\/}.  \publ{New York: Dover}.

\bibitem[Coleman \& Vassilicos(2009)]{coleman_unified_2009}
{\sc \au{Coleman, S.~W.} \& \au{Vassilicos, J.~C.}} \yr{2009}  \at{A unified
  sweep-stick mechanism to explain particle clustering in two- and
  three-dimensional homogeneous, isotropic turbulence}.  \jt{Phys. Fluids}
  \bvol{21}~(11),  \pg{113301}.

\bibitem[Darabi {\em et~al.\/}(2019)Darabi, Koleini, Deglon, Rezai \&
  Abdollahy]{darabi_investigation_2019}
{\sc \au{Darabi, H.}, \au{Koleini, S. M.~J.}, \au{Deglon, D.}, \au{Rezai, B.}
  \& \au{Abdollahy, M.}} \yr{2019}  \at{Investigation of bubble-particle
  interactions in a mechanical flotation cell, part 1: {Collision} frequencies
  and efficiencies}.  \jt{Miner. Eng.}  \bvol{134},  \pg{54--64}.

\bibitem[Eswaran \& Pope(1988)]{eswaran_examination_1988}
{\sc \au{Eswaran, V.} \& \au{Pope, S.~B.}} \yr{1988}  \at{An examination of
  forcing in direct numerical simulations of turbulence}.  \jt{Comput. Fluids}
  \bvol{16}~(3),  \pg{257--278}.

\bibitem[Falkovich {\em et~al.\/}(2002)Falkovich, Fouxon \&
  Stepanov]{falkovich_acceleration_2002}
{\sc \au{Falkovich, G.}, \au{Fouxon, A.} \& \au{Stepanov, M.~G.}} \yr{2002}
  \at{Acceleration of rain initiation by cloud turbulence}.  \jt{Nature}
  \bvol{419}~(6903),  \pg{151--154}.

\bibitem[Falkovich \& Pumir(2007)]{falkovich_sling_2007}
{\sc \au{Falkovich, G.} \& \au{Pumir, A.}} \yr{2007}  \at{Sling effect in
  collisions of water droplets in turbulent clouds}.  \jt{J. Atmos. Sci.}
  \bvol{64}~(12),  \pg{4497--4505}.

\bibitem[Fayed \& Ragab(2013)]{fayed_direct_2013}
{\sc \au{Fayed, H.~E.} \& \au{Ragab, S.~A.}} \yr{2013}  \at{Direct numerical
  simulation of particles--bubbles collisions kernel in homogeneous isotropic
  turbulence}.  \jt{J. Comput. Multph. Flows}  \bvol{5}~(3),  \pg{167--188}.

\bibitem[Fouxon(2012)]{fouxon_distribution_2012}
{\sc \au{Fouxon, I.}} \yr{2012}  \at{Distribution of particles and bubbles in
  turbulence at a small {Stokes} number}.  \jt{Phys. Rev. Lett.}
  \bvol{108}~(13),  \pg{134502}.

\bibitem[Goto \& Vassilicos(2006)]{goto_self-similar_2006}
{\sc \au{Goto, S.} \& \au{Vassilicos, J.~C.}} \yr{2006}  \at{Self-similar
  clustering of inertial particles and zero-acceleration points in fully
  developed two-dimensional turbulence}.  \jt{Phys. Fluids}  \bvol{18}~(11),
  \pg{115103}.

\bibitem[Goto \& Vassilicos(2008)]{goto_sweep-stick_2008}
{\sc \au{Goto, S.} \& \au{Vassilicos, J.~C.}} \yr{2008}  \at{Sweep-stick
  mechanism of heavy particle clustering in fluid turbulence}.  \jt{Phys. Rev.
  Lett.}  \bvol{100}~(5),  \pg{054503}.

\bibitem[Hassanzadeh {\em et~al.\/}(2018)Hassanzadeh, Firouzi, Albijanic \&
  Celik]{hassanzadeh_review_2018}
{\sc \au{Hassanzadeh, A.}, \au{Firouzi, M.}, \au{Albijanic, B.} \& \au{Celik,
  M.~S.}} \yr{2018}  \at{A review on determination of particle–bubble
  encounter using analytical, experimental and numerical methods}.  \jt{Miner.
  Eng.}  \bvol{122},  \pg{296--311}.

\bibitem[van Hinsberg {\em et~al.\/}(2017)van Hinsberg, Clercx \&
  Toschi]{van_hinsberg_enhanced_2017}
{\sc \au{van Hinsberg, M. A.~T.}, \au{Clercx, H. J.~H.} \& \au{Toschi, F.}}
  \yr{2017}  \at{Enhanced settling of nonheavy inertial particles in
  homogeneous isotropic turbulence: {The} role of the pressure gradient and the
  {Basset} history force}.  \jt{Phys. Rev. E}  \bvol{95}~(2),  \pg{023106}.

\bibitem[Homann \& Bec(2010)]{homann_finite-size_2010}
{\sc \au{Homann, H.} \& \au{Bec, J.}} \yr{2010}  \at{Finite-size effects in the
  dynamics of neutrally buoyant particles in turbulent flow}.  \jt{J. Fluid
  Mech.}  \bvol{651},  \pg{81--91}.

\bibitem[Huang {\em et~al.\/}(2012)Huang, Legendre \&
  Guiraud]{huang_effect_2012}
{\sc \au{Huang, Z.}, \au{Legendre, D.} \& \au{Guiraud, P.}} \yr{2012}
  \at{Effect of interface contamination on particle–bubble collision}.
  \jt{Chem. Eng. Sci.}  \bvol{68}~(1),  \pg{1--18}.

\bibitem[Ijzermans {\em et~al.\/}(2010)Ijzermans, Meneguz \&
  Reeks]{ijzermans_segregation_2010}
{\sc \au{Ijzermans, R. H.~A.}, \au{Meneguz, E.} \& \au{Reeks, M.~W.}} \yr{2010}
   \at{Segregation of particles in incompressible random flows: singularities,
  intermittency and random uncorrelated motion}.  \jt{J. Fluid Mech.}
  \bvol{653},  \pg{99--136}.

\bibitem[Ireland {\em et~al.\/}(2016)Ireland, Bragg \&
  Collins]{ireland_effect_2016}
{\sc \au{Ireland, P.~J.}, \au{Bragg, A.~D.} \& \au{Collins, L.~R.}} \yr{2016}
  \at{The effect of {Reynolds} number on inertial particle dynamics in
  isotropic turbulence. {Part} 1. {Simulations} without gravitational effects}.
   \jt{J. Fluid Mech.}  \bvol{796},  \pg{617--658}.

\bibitem[Jiménez {\em et~al.\/}(1993)Jiménez, Wray, Saffman \&
  Rogallo]{jimenez_structure_1993}
{\sc \au{Jiménez, J.}, \au{Wray, A.~A.}, \au{Saffman, P.~G.} \& \au{Rogallo,
  R.~S.}} \yr{1993}  \at{The structure of intense vorticity in isotropic
  turbulence}.  \jt{J. Fluid Mech.}  \bvol{255},  \pg{65--90}.

\bibitem[Kolmogorov(1941)]{kolmogorov_local_1941}
{\sc \au{Kolmogorov, A.~N.}} \yr{1941}  \at{The local structure of turbulence
  in incompressible viscous fluid for very large {Reynolds} number}.  \jt{Dokl.
  Akad. Nauk USSR}  \bvol{30},  \pg{299--303}.

\bibitem[Kolmogorov(1963)]{kolmogorov_approximation_1963}
{\sc \au{Kolmogorov, A.~N.}} \yr{1963}  \at{On the approximation of
  distributions of sums of independent summands by infinitely divisible
  distributions}.  \jt{Sankhyā: Indian J. Stat. A}  \bvol{25}~(2),
  \pg{159--174}.

\bibitem[Kostoglou {\em et~al.\/}(2020{\natexlab{{\em a\/}}})Kostoglou,
  Karapantsios \& Evgenidis]{kostoglou_generalized_2020}
{\sc \au{Kostoglou, M.}, \au{Karapantsios, T.~D.} \& \au{Evgenidis, S.}}
  \yr{2020{\natexlab{{\em a\/}}}}  \at{On a generalized framework for turbulent
  collision frequency models in flotation: {The} road from past inconsistencies
  to a concise algebraic expression for fine particles}.  \jt{Adv. Colloid
  Interface Sci.}  \bvol{284},  \pg{102270}.

\bibitem[Kostoglou {\em et~al.\/}(2020{\natexlab{{\em b\/}}})Kostoglou,
  Karapantsios \& Oikonomidou]{kostoglou_critical_2020}
{\sc \au{Kostoglou, M.}, \au{Karapantsios, T.~D.} \& \au{Oikonomidou, O.}}
  \yr{2020{\natexlab{{\em b\/}}}}  \at{A critical review on turbulent collision
  frequency/efficiency models in flotation: {Unravelling} the path from general
  coagulation to flotation}.  \jt{Adv. Colloid Interface Sci.}  \bvol{279},
  \pg{102158}.

\bibitem[Kruis \& Kusters(1997)]{kruis_collision_1997}
{\sc \au{Kruis, F.~E.} \& \au{Kusters, K.~A.}} \yr{1997}  \at{The collision
  rate of particles in turbulent flow}.  \jt{Chem. Eng. Comm.}  \bvol{158}~(1),
   \pg{201--230}.

\bibitem[Li {\em et~al.\/}(2021)Li, Abraham, Guala \& Hong]{li_evidence_2021}
{\sc \au{Li, J.}, \au{Abraham, A.}, \au{Guala, M.} \& \au{Hong, J.}} \yr{2021}
  \at{Evidence of preferential sweeping during snow settling in atmospheric
  turbulence}.  \jt{J. Fluid Mech.}  \bvol{928},  \pg{A8}.

\bibitem[Liepe \& Möckel(1976)]{liepe_untersuchungen_1976}
{\sc \au{Liepe, F.} \& \au{Möckel, H.-O.}} \yr{1976}  \at{Untersuchungen zum
  stoffvereinigen in flüssiger phase}.  \jt{Chem. Techn.}  \bvol{28}~(4),
  \pg{205--209}.

\bibitem[Maxey(1987)]{maxey_gravitational_1987}
{\sc \au{Maxey, M.~R.}} \yr{1987}  \at{The gravitational settling of aerosol
  particles in homogeneous turbulence and random flow fields}.  \jt{J. Fluid
  Mech.}  \bvol{174},  \pg{441--465}.

\bibitem[Maxey \& Riley(1983)]{maxey_equation_1983}
{\sc \au{Maxey, M.~R.} \& \au{Riley, J.~J.}} \yr{1983}  \at{Equation of motion
  for a small rigid sphere in a nonuniform flow}.  \jt{Phys. Fluids}
  \bvol{26}~(4),  \pg{883--889}.

\bibitem[Mazzitelli {\em et~al.\/}(2003)Mazzitelli, Lohse \&
  Toschi]{mazzitelli_effect_2003}
{\sc \au{Mazzitelli, I.~M.}, \au{Lohse, D.} \& \au{Toschi, F.}} \yr{2003}
  \at{The effect of microbubbles on developed turbulence}.  \jt{Phys. Fluids}
  \bvol{15}~(1),  \pg{L5}.

\bibitem[Mehlig {\em et~al.\/}(2007)Mehlig, Uski \&
  Wilkinson]{mehlig_colliding_2007}
{\sc \au{Mehlig, B.}, \au{Uski, V.} \& \au{Wilkinson, M.}} \yr{2007}
  \at{Colliding particles in highly turbulent flows}.  \jt{Phys. Fluids}
  \bvol{19}~(9),  \pg{098107}.

\bibitem[Miettinen {\em et~al.\/}(2010)Miettinen, Ralston \&
  Fornasiero]{miettinen_limits_2010}
{\sc \au{Miettinen, T.}, \au{Ralston, J.} \& \au{Fornasiero, D.}} \yr{2010}
  \at{The limits of fine particle flotation}.  \jt{Miner. Eng.}  \bvol{23}~(5),
   \pg{420--437}.

\bibitem[Monchaux {\em et~al.\/}(2010)Monchaux, Bourgoin \&
  Cartellier]{monchaux_preferential_2010}
{\sc \au{Monchaux, R.}, \au{Bourgoin, M.} \& \au{Cartellier, A.}} \yr{2010}
  \at{Preferential concentration of heavy particles: {A} {Voronoï} analysis}.
  \jt{Phys. Fluids}  \bvol{22}~(10),  \pg{103304}.

\bibitem[Ngo-Cong {\em et~al.\/}(2018)Ngo-Cong, Nguyen \&
  Tran-Cong]{ngo-cong_isotropic_2018}
{\sc \au{Ngo-Cong, D.}, \au{Nguyen, A.~V.} \& \au{Tran-Cong, T.}} \yr{2018}
  \at{Isotropic turbulence surpasses gravity in affecting bubble-particle
  collision interaction in flotation}.  \jt{Miner. Eng.}  \bvol{122},
  \pg{165--175}.

\bibitem[Nguyen {\em et~al.\/}(2016)Nguyen, An-Vo, Tran-Cong \&
  Evans]{nguyen2016}
{\sc \au{Nguyen, A.~V.}, \au{An-Vo, D.-A.}, \au{Tran-Cong, T.} \& \au{Evans,
  G.~M.}} \yr{2016}  \at{A review of stochastic description of the turbulence
  effect on bubble-particle interactions in flotation}.  \jt{Int. J. Miner.
  Process.}  \bvol{156},  \pg{75--86}.

\bibitem[Nguyen {\em et~al.\/}(2006)Nguyen, George \&
  Jameson]{nguyen_demonstration_2006}
{\sc \au{Nguyen, A.~V.}, \au{George, P.} \& \au{Jameson, G.~J.}} \yr{2006}
  \at{Demonstration of a minimum in the recovery of nanoparticles by flotation:
  {Theory} and experiment}.  \jt{Chem. Eng. Sci.}  \bvol{61}~(8),
  \pg{2494--2509}.

\bibitem[Nguyen \& Schulze(2004)]{nguyen_colloidal_2004}
{\sc \au{Nguyen, A.~V.} \& \au{Schulze, H.~J.}} \yr{2004} {\em Colloidal
  science of flotation\/}, 1st edn.,  \st{Surfactant sciences},  \vol{vol.
  118}.  \publ{Boca Raton: CRC Press}.

\bibitem[Obligado {\em et~al.\/}(2014)Obligado, Teitelbaum, Cartellier, Mininni
  \& Bourgoin]{obligado_preferential_2014}
{\sc \au{Obligado, M.}, \au{Teitelbaum, T.}, \au{Cartellier, A.}, \au{Mininni,
  P.} \& \au{Bourgoin, M.}} \yr{2014}  \at{Preferential concentration of heavy
  particles in turbulence}.  \jt{J. Turbul.}  \bvol{15}~(5),  \pg{293--310}.

\bibitem[Ostilla-Monico {\em et~al.\/}(2015)Ostilla-Monico, Yang, van~der Poel,
  Lohse \& Verzicco]{ostilla-monico_multiple-resolution_2015}
{\sc \au{Ostilla-Monico, R.}, \au{Yang, Y.}, \au{van~der Poel, E.~P.},
  \au{Lohse, D.} \& \au{Verzicco, R.}} \yr{2015}  \at{A multiple-resolution
  strategy for direct numerical simulation of scalar turbulence}.  \jt{J.
  Comput. Phys.}  \bvol{301},  \pg{308--321}.

\bibitem[Petersen {\em et~al.\/}(2019)Petersen, Baker \&
  Coletti]{petersen_experimental_2019}
{\sc \au{Petersen, A.~J.}, \au{Baker, L.} \& \au{Coletti, F.}} \yr{2019}
  \at{Experimental study of inertial particles clustering and settling in
  homogeneous turbulence}.  \jt{J. Fluid Mech.}  \bvol{864},  \pg{925--970}.

\bibitem[van~der Poel {\em et~al.\/}(2015)van~der Poel, Ostilla-Mónico,
  Donners \& Verzicco]{van_der_poel_pencil_2015}
{\sc \au{van~der Poel, E.~P.}, \au{Ostilla-Mónico, R.}, \au{Donners, J.} \&
  \au{Verzicco, R.}} \yr{2015}  \at{A pencil distributed finite difference code
  for strongly turbulent wall-bounded flows}.  \jt{Comput. Fluids}  \bvol{116},
   \pg{10--16}.

\bibitem[Pumir \& Wilkinson(2016)]{pumir_collisional_2016}
{\sc \au{Pumir, A.} \& \au{Wilkinson, M.}} \yr{2016}  \at{Collisional
  aggregation due to turbulence}.  \jt{Annu. Rev. Condens. Matter Phys.}
  \bvol{7},  \pg{141--170}.

\bibitem[Rogich \& Matos(2008)]{rogich_global_2008}
{\sc \au{Rogich, D.~G.} \& \au{Matos, G.~R.}} \yr{2008}  \bt{The global flows
  of metals and minerals}. Open-{File} {Report} 2008-1355.  \org{U.S.
  Geological Survey}, Reston.

\bibitem[Saffman \& Turner(1956)]{saffman_collision_1956}
{\sc \au{Saffman, P.~G.} \& \au{Turner, J.~S.}} \yr{1956}  \at{On the collision
  of drops in turbulent clouds}.  \jt{J. Fluid Mech.}  \bvol{1}~(1),
  \pg{16--30}.

\bibitem[Saw(2008)]{saw_studies_2008}
{\sc \au{Saw, E.~W.}} \yr{2008}  \at{Studies of spatial clustering of inertial
  particles in turbulence}. {PhD} thesis, Michigan Technological University.

\bibitem[Schiller \& Naumann(1933)]{schiller_uber_1933}
{\sc \au{Schiller, L.} \& \au{Naumann, A.Z.}} \yr{1933}  \at{Über die
  grundlegenden berechnungen bei der schwerkraftaufbereitung}.  \jt{Z. Ver.
  Dtsch. Ing.,}  \bvol{77}~(12),  \pg{318--320}.

\bibitem[Schubert(1999)]{schubert_turbulence-controlled_1999}
{\sc \au{Schubert, H.}} \yr{1999}  \at{On the turbulence-controlled
  microprocesses in flotation machines}.  \jt{Int. J. Miner. Process.}
  \bvol{56}~(1),  \pg{257--276}.

\bibitem[Spandan {\em et~al.\/}(2020)Spandan, Putt, Ostilla-Mónico \&
  Lee]{spandan_fluctuation-induced_2020}
{\sc \au{Spandan, V.}, \au{Putt, D.}, \au{Ostilla-Mónico, R.} \& \au{Lee,
  A.~A.}} \yr{2020}  \at{Fluctuation-induced force in homogeneous isotropic
  turbulence}.  \jt{Sci. Adv.}  \bvol{6}~(14),  \pg{eaba0461}.

\bibitem[Sundaram \& Collins(1996)]{sundaram_numerical_1996}
{\sc \au{Sundaram, S.} \& \au{Collins, L.~R.}} \yr{1996}  \at{Numerical
  considerations in simulating a turbulent suspension of finite-volume
  particles}.  \jt{J. Comput. Phys.}  \bvol{124}~(2),  \pg{337--350}.

\bibitem[Sundaram \& Collins(1997)]{sundaram_collision_1997}
{\sc \au{Sundaram, S.} \& \au{Collins, L.~R.}} \yr{1997}  \at{Collision
  statistics in an isotropic particle-laden turbulent suspension. {Part} 1.
  {Direct} numerical simulations}.  \jt{J. Fluid Mech.}  \bvol{335},
  \pg{75--109}.

\bibitem[Tchen(1947)]{tchen_mean_1947}
{\sc \au{Tchen, C.-M.}} \yr{1947}  \at{Mean value and correlation problems
  connected with the motion of small particles suspended in a turbulent fluid}.
  {PhD} thesis, Technische Hogeschool Delft, Delft.

\bibitem[Verzicco \& Orlandi(1996)]{verzicco_finite-difference_1996}
{\sc \au{Verzicco, R.} \& \au{Orlandi, P.}} \yr{1996}  \at{A finite-difference
  scheme for three-dimensional incompressible flows in cylindrical
  coordinates}.  \jt{J. Comput. Phys.}  \bvol{123}~(2),  \pg{402--414}.

\bibitem[Völk {\em et~al.\/}(1980)Völk, Jones, Morfill \&
  Röser]{volk_collisions_1980}
{\sc \au{Völk, H.~J.}, \au{Jones, F.~C.}, \au{Morfill, G.~E.} \& \au{Röser,
  S.}} \yr{1980}  \at{Collisions between grains in a turbulent gas}.
  \jt{Astron. Astrophys.}  \bvol{85}~(3),  \pg{316--325}.

\bibitem[Voßkuhle {\em et~al.\/}(2014)Voßkuhle, Pumir, Lévêque \&
  Wilkinson]{voskuhle_prevalence_2014}
{\sc \au{Voßkuhle, M.}, \au{Pumir, A.}, \au{Lévêque, E.} \& \au{Wilkinson,
  M.}} \yr{2014}  \at{Prevalence of the sling effect for enhancing collision
  rates in turbulent suspensions}.  \jt{J. Fluid Mech.}  \bvol{749},
  \pg{841--852}.

\bibitem[Wan {\em et~al.\/}(2020)Wan, Yi, Wang, Sun, Chen \&
  Wang]{wan_study_2020}
{\sc \au{Wan, D.}, \au{Yi, X.}, \au{Wang, L.-P.}, \au{Sun, X.}, \au{Chen, S.}
  \& \au{Wang, G.}} \yr{2020}  \at{Study of collisions between particles and
  unloaded bubbles with point-particle model embedded in the direct numerical
  simulation of turbulent flows}.  \jt{Miner. Eng.}  \bvol{146},  \pg{106137}.

\bibitem[Wang {\em et~al.\/}(2005)Wang, Ayala \& Xue]{wang_reconciling_2005}
{\sc \au{Wang, L.-P.}, \au{Ayala, O.} \& \au{Xue, Y.}} \yr{2005}
  \at{Reconciling the cylindrical formulation with the spherical formulation in
  the kinematic descriptions of collision kernel}.  \jt{Phys. Fluids}
  \bvol{17}~(6),  \pg{067103}.

\bibitem[Wang {\em et~al.\/}(1998)Wang, Wexler \& Zhou]{wang_collision_1998}
{\sc \au{Wang, L.-P.}, \au{Wexler, A.~S.} \& \au{Zhou, Y.}} \yr{1998}  \at{On
  the collision rate of small particles in isotropic turbulence. {I}.
  {Zero}-inertia case}.  \jt{Phys. Fluids}  \bvol{10}~(1),  \pg{266--276}.

\bibitem[Wang {\em et~al.\/}(2020)Wang, Wan, Yang, Wang \&
  Chen]{wang_reynolds_2020}
{\sc \au{Wang, X.}, \au{Wan, M.}, \au{Yang, Y.}, \au{Wang, L.-P.} \& \au{Chen,
  S.}} \yr{2020}  \at{Reynolds number dependence of heavy particles clustering
  in homogeneous isotropic turbulence}.  \jt{Phys. Rev. Fluids}  \bvol{5}~(12),
   \pg{124603}.

\bibitem[Wilkinson {\em et~al.\/}(2006)Wilkinson, Mehlig \&
  Bezuglyy]{wilkinson_caustic_2006}
{\sc \au{Wilkinson, M.}, \au{Mehlig, B.} \& \au{Bezuglyy, V.}} \yr{2006}
  \at{Caustic activation of rain showers}.  \jt{Phys. Rev. Lett.}
  \bvol{97}~(4),  \pg{048501}.

\bibitem[Williams \& Crane(1983)]{williams_particle_1983}
{\sc \au{Williams, J. J.~E.} \& \au{Crane, R.~I.}} \yr{1983}  \at{Particle
  collision rate in turbulent flow}.  \jt{Int. J. Multiphase Flow}
  \bvol{9}~(4),  \pg{421--435}.

\bibitem[{World Bank Group}(2017)]{world_bank_group_growing_2017}
{\sc \au{{World Bank Group}}} \yr{2017} {\em The growing role of minerals and
  metals for a low carbon future\/}.  \publ{World Bank}.

\bibitem[Yuu(1984)]{yuu_collision_1984}
{\sc \au{Yuu, S.}} \yr{1984}  \at{Collision rate of small particles in a
  homogeneous and isotropic turbulence}.  \jt{AIChE J.}  \bvol{30}~(5),
  \pg{802--807}.

\bibitem[Zaichik \& Alipchenkov(2009)]{zaichik_statistical_2009}
{\sc \au{Zaichik, L.~I} \& \au{Alipchenkov, V.~M}} \yr{2009}  \at{Statistical
  models for predicting pair dispersion and particle clustering in isotropic
  turbulence and their applications}.  \jt{New J. Phys.}  \bvol{11}~(10),
  \pg{103018}.

\bibitem[Zaichik {\em et~al.\/}(2010)Zaichik, Simonin \&
  Alipchenkov]{zaichik_turbulent_2010}
{\sc \au{Zaichik, L.~I.}, \au{Simonin, O.} \& \au{Alipchenkov, V.~M.}}
  \yr{2010}  \at{Turbulent collision rates of arbitrary-density particles}.
  \jt{Int. J. Heat Mass Transf.}  \bvol{53}~(9),  \pg{1613--1620}.

\bibitem[Zhou {\em et~al.\/}(2001)Zhou, Wexler \& Wang]{zhou_modelling_2001}
{\sc \au{Zhou, Y.}, \au{Wexler, A.~S.} \& \au{Wang, L.-P.}} \yr{2001}
  \at{Modelling turbulent collision of bidisperse inertial particles}.  \jt{J.
  Fluid Mech.}  \bvol{433},  \pg{77--104}.

\end{thebibliography}
	
\end{document}